\documentclass[traditabstract]{aa}
\usepackage{txfonts}
\usepackage{graphicx}
\usepackage{graphicx}
\usepackage{natbib}
\bibpunct{(}{)}{;}{a}{}{,}
\usepackage{longtable}

\long\def\symbolfootnote[1]{\begingroup\def\thefootnote{\fnsymbol{footnote}}\footnote[1]\endgroup}

\begin{document}

\title{Galaxy protocluster candidates at $1.6<z\lesssim2$\thanks{Based on observations obtained at the European Southern Observatory using the Very Large Telescope on Cerro Paranal through ESO programs 081.A-0673(A) and 083.A-0231(A).}}

\titlerunning{Galaxy protocluster candidates around radio galaxies at $1.6<z\lesssim2$}

\author{Audrey Galametz\inst{1,2}
\and Jo\"{e}l Vernet\inst{1}
\and Carlos De Breuck\inst{1}
\and Nina Hatch\inst{3,4}
\and George Miley\inst{3} \and \\
Tadayuki Kodama\inst{5}
\and Jaron Kurk\inst{6} 
\and Roderik Overzier\inst{7}
\and Alessandro Rettura\inst{8} \and \\
Huub R\"{o}ttgering\inst{3} 
\and Nick Seymour\inst{9}
\and Bram Venemans\inst{1}
\and Andrew Zirm\inst{10}
}

\institute{European Southern Observatory, Karl-Schwarzschild-Str. 2, D-85748 Garching, Germany [e-mail: {\tt agalamet@eso.org}]
\and Observatoire Astronomique de Strasbourg, 11 rue de l$'$Universit\'e, 67000 Strasbourg, France
\and Leiden Observatory, University of Leiden, P.B. 9513, Leiden 2300 RA, The Netherlands
\and University of Nottingham, School of Physics and Astronomy, Nottingham NG7 2RD
\and National Astronomical Observatory of Japan, Mitaka, Tokyo 181-8588, Japan
\and Max-Planck-Institut fŸr extraterrestrische Physik (MPE), Giessenbachstr.1, D-85748 Garching, Germany
\and Max-Planck-Institut fŸr Astrophysik (MPA), D-85748 Garching, Germany 
\and Department of Physics and Astronomy, University of California, Riverside, CA 92521, USA 
\and Mullard Space Science Laboratory, UCL, Holmbury St Mary, Dorking, Surrey, RH5 6NT
\and Dark Cosmology Centre, Niels Bohr Institute, University of Copenhagen, Juliane Mariesvej 30, DK-2100 Copenhagen, Denmark 
}

\abstract{We present a study of protoclusters associated with high redshift radio galaxies. We 
imaged MRC~1017-220 ($z=1.77$) and MRC~0156-252 ($z=2.02$) using the near-infrared 
wide-field ($7.5\arcmin \times 7.5\arcmin$) imager VLT/HAWK-I in the $Y$, $H$ and $Ks$ 
bands. We present the first deep $Y$-band galaxy number counts within a large area ($\sim$200 arcmin$^2$). 
We then develop a purely near-infrared colour selection 
technique to isolate galaxies at $1.6<z<3$ that may be associated 
with the two targets, dividing them into {\it (i)} red passively evolving or dusty star-forming 
galaxies or {\it (ii)} blue/star-formation dominated galaxies with little or no dust. Both targeted 
fields show an excess of star-forming galaxies with respect to control fields. No clear 
overdensity of red galaxies is detected in the surroundings of MRC~1017-220 although the 
spatial distribution of the red galaxies resembles a filament-like structure within which the radio galaxy is embedded. 
In contrast, a significant overdensity of red galaxies is detected in the field of MRC~0156-252,
ranging from a factor of $\sim2-3$ times the field density at large scales ($2.5$~Mpc, angular distance) up 
to a factor of $\sim3-4$ times the field density within a $1$~Mpc radius of the radio galaxy. Half of these red galaxies have 
colours consistent with red sequence models at $z\sim2$, with a large fraction being bright 
($Ks<21.5$, i.e.~massive). In addition, we also find a small group of galaxies 
within 5\arcsec\ of MRC~0156-252 suggesting that the radio galaxy has multiple companions within 
$\sim50$~kpc. We conclude that the field of MRC0156-252 shows many remarkable 
similarities with the well-studied protocluster surrounding PKS1138-262 
($z$=2.16) suggesting that MRC~0156-252 is associated with a galaxy protocluster at $z\sim2$.}


\maketitle

\section{Introduction}

It is well established that the evolution of galaxies strongly depends on  
environment. In the nearby Universe, the highest density regions (e.g., the 
cores of local galaxy clusters) are dominated by red, early-type galaxies with 
the fraction of blue galaxies with on-going star-formation significantly smaller 
than lower density regions \citep{Dressler1980, Tanaka2005, Postman2005, Balogh2007, Poggianti2009}. 
The spectra of these red passively evolving galaxies show a characteristic 
break at $4000\AA$ --- i.e., the light from old stellar populations is more prominent 
than that emitted by younger stars. These galaxies lie on a 
tight red sequence in colour-magnitude diagrams. Studies 
have shown that this red sequence is already well populated in clusters of galaxies 
out to high redshifts \citep[$z\sim1.5$;][]{Stanford2005, Stanford2006, Mei2006, 
Mei2009, Lidman2008, Kurk2009}. To understand when this red sequence appeared, 
i.e.,~when the segregation between passive and star-forming galaxies occurred 
in clusters, one needs to study overdensities at higher redshifts. 

However, the number of known galaxy clusters at high redshifts ($z>1$) is small.
Indeed, searching for higher redshift clusters rapidly becomes challenging using classical
detection methods such as using red-sequence algorithms aimed at identifying red 
galaxy overdensities \citep{Gladders2000, Andreon2008} or detecting the extended 
X-ray emission from the intracluster medium \citep{Stanford2006, Rosati2004, Rosati2009}. 
Until recently, the two
highest redshift clusters were discovered through X-rays: 
XMMXCS~J2215.9-1738 at $z=1.457$ \citep[with $17$ spectroscopically confirmed members
within the cluster virial radius;][]{Stanford2006, Hilton2007} and XMMU~J2235.3-2557 at 
$z=1.39$ \citep[with $34$ spectroscopically
confirmed members;][]{Rosati2009}. Galaxies in the core of this second cluster already lie on a 
well defined and tight red sequence \citep{Lidman2008}. These X-ray selected clusters were 
however recently supersceded by the discovery of a galaxy cluster at $z=1.62$, ClG~J0218-0510, 
using photometric redshifts. It is located in the Subaru/XMM-Newton deep 
field and has $15$ confirmed members to date \citep{Tanaka2010, Papovich2010}. 

One of the most efficient methods to search for galaxy clusters at even higher 
redshifts ($z>1.5$) is to look in the vicinity of high-redshift radio galaxies 
(HzRGs hereafter). These galaxies are among the most massive galaxies in the Universe 
\citep[M$>10^{11}$M$_{\odot}$;][]{Rocca2004, Seymour2007} and are good tracers
of high density regions in the early universe \citep[see][for a review on HzRGs and their 
surroundings]{Miley2008}. Narrow line emitter surveys of HzRGs at $2<z<5$ show 
that they are often located in overdense regions, designated as 
`protoclusters' that are likely to be the progenitors of the present 
day massive groups and clusters \citep{Venemans2002, Venemans2005,Venemans2007, 
Pentericci2000, Kurk2004A}. The narrow line emitters are, however, only a low-mass subset 
of the general population of UV-selected, star-forming galaxies \citep{Miley2004, Overzier2008} 
and likely do not represent the majority of the total stellar mass.

Studies have tried to identify more massive cluster galaxies associated with 
these radio-loud sources by looking for the red passively evolving galaxies
that may be populating the cores of high redshift clusters.
\citet{Best2003} observed, for example, the environment of powerful radio-loud 
sources at $z\sim1.6$ and found overdensities of red galaxies ($R-K>4$) 
on two scales around the AGN: a pronounced central peak (within $150$~kpc) and 
weaker excesses between $1$ and $1.5$~Mpc radius. More recently, \citet{Galametz2009B} 
studied the environment of 7C~1756+6520, a radio galaxy at $z=1.416$, and found 
an excess of red sources (passive, early-type galaxy candidates 
at $z>1.4$) within $2$~Mpc of the HzRG. A galaxy cluster associated with 7C~1756+6520
has since been spectroscopically confirmed \citep{Galametz2010}. At higher redshifts ($z>2$), studies 
have searched for red evolved galaxies by bracketing the redshifted $4000\AA$~break  
with near-infrared filters. \citet{Kajisawa2006} explored the environments 
of six HzRGs at $z\sim2.5$ and isolated the evolved galaxy population at $z>2$ using 
purely near-infrared ($JHKs$) colour cuts. \citet{Kodama2007} used the same
near-infrared criteria to select protocluster member candidates in the field of HzRGs at $2<z\lesssim3$ 
\citep[see also][~for a near-infrared study of a forming red sequence in a protocluster at $z=2.16$]{Zirm2008}.
They both found that some of their targeted fields contained overdensities by a factor of $2-3$ 
compared to blank fields. Recent spectroscopic follow-up of these red sources 
has been conducted in a couple of these protocluster fields by \citet{Doherty2010}. They confirm two red
galaxies associated with PKS~1138-262 at $z=2.16$,  
a dusty star-forming galaxy and an evolved galaxy with little ongoing star formation.
These HzRG companions have an estimated mass of $4-6 \times 10^{11}$M$_{\odot}$. 
\citet{Doherty2010} also confirm that a pure near-infrared criterion is efficient at selecting 
high redshift galaxies --- e.g.~$56$\% of their $JHK$-selected galaxies with spectroscopic redshift
fall at $2.3<z<3.1$. However, the low success rate of their spectroscopic campaign confirms
however the challenge in deriving redshifts for passively evolving galaxies whose spectra
do not show prominent and easily identifiable emission lines.

Building on these previous studies of individual HzRG fields, we defined a first 
uniformly selected sample of the most powerful radio galaxies in the pivotal redshift 
range $1.7<z<2.6$ where we expect the cluster galaxies to start settling on the red 
sequence. Our method is to select potential cluster members (in particular, evolved 
passive galaxies) using near-infrared colour cuts. We observed our sample with the High 
Acuity Wide field K-band Imager \citep[HAWK-I;][]{Pirard2004, Casali2006, 
Kissler2008} on the Very Large Telescope (VLT) in 2008 and 2009 in a set of three filters 
 ($YHKs$ or $JHKs$ depending on the redshift of the targeted HzRG).
This paper reports the results on our two lowest redshift targets. A companion paper 
presents results on the higher redshift targets (Hatch et al.,~2010).

We design a news near-infrared criterion to isolate galaxies at $1.6<z<3$. We apply this 
selection technique to study the galaxy population in the vicinity of the lowest redshift HzRGs 
of the HAWK-I sample: MRC~1017-220 ($z=1.77$) and MRC~0156-252 ($z=2.02$). We first 
describe in \S2 the multi-wavelength data available for the targets and four control fields (including two sub-fields 
of GOODS South). The extraction of the source catalogues and galaxy number counts are described in \S3 and \S4. 
Section 5 presents the colour-colour selection technique we developed to isolate cluster member candidates at $z>1.6$.
The properties of these cluster candidates, such as overdensities and spatial distribution 
are detailed in \S6. Section 7 summarizes our results.

We assume a $\Lambda$CDM cosmology with $H_0 = 70$ 
km s$^{-1}$ Mpc$^{-1}$, $\Omega_m = 0.3$ and $\Omega_{\Lambda} = 0.7$. All magnitudes 
are expressed in the AB photometric system unless stated otherwise. 

\section{The data}

\subsection{The targets}

We present a study of the two lowest redshifts targets of the HAWK-I sample: 
MRC~1017-220 and MRC~0156-252, observed with HAWK-I in the $Y$, $H$ and $Ks$ bands.

MRC~1017-220 ($z=1.768$; R.A.: 10:19:49.05, Dec.: -22:19:58.03, J2000, 
$L_{3GHz}=10^{28.11}$~W Hz$^{-1}$) is our 
lowest redshift targeted HzRG. We note that this HzRG is a broad line radio galaxy \citep{Kapahi1998} 
and is unresolved both in near-infrared and radio \citep{Pentericci2001}. Investigating extremely red 
objects (EROs; $R-K>6$) around high-$z$ AGN, \citet{Cimatti2000} found an excess of EROs in the 
close vicinity of MRC~1017-220 compared to the field. 

We also targeted the field around MRC~0156-252 ($z=2.016$; R.A.: 01:58:33.63, Dec.: -24:59:31.10,
$L_{3GHz}=10^{27.79}$~W Hz$^{-1}$).
This HzRG has been reported to be a quasar obscured by dust \citep{Eales1996,McCarthy1992}. 

\subsection{Observations and data reduction}

\subsubsection{New VLT/HAWK-I data}

\begin{table*}
\caption{HAWK-I observations}
\centering
\begin{tabular}{l l l l c c l}
Field		&	R.A.$^{\mathrm{a}}$		&	Dec.$^{\mathrm{a}}$		&	Band	&	Exp.Time	&	Seeing	&	Limit		\\
		&	J2000				&	J2000				&			&	min		&	arcsec	&	$2\sigma$ ($3\sigma$)	\\
\hline
MRC~1017-220	&	10:19:54.18		&	-22:18:27.17	&	$Y$	&	$122$	&	$0.80$	&	$25.91$ ($25.47$)	\\	
				&					&				&	$H$	&	$53$		&	$0.63$	&	$24.89$ ($24.45$)	\\
				&					&				&	$Ks$	&	$33$		&	$0.55$	&	$24.36$ ($23.92$)	\\
MRC~0156-252	&	01:58:39.16		&	-25:00:51.01	&	$Y$	&	$200$	&	$0.54$	&	$26.42$ ($25.98$)	\\
				&					&				&	$H$	&	$47$		&	$0.50$	&	$24.91$ ($24.47$)	\\
				&					&				&	$Ks$	&	$33$		&	$0.56$	&	$24.42$ ($23.98$)	\\
CF1				&	11:39:59.66		&	-11:24:29.50	&	$Y$	&	$122$	&	$0.58$	&	$25.80$ ($25.36$)	\\
				&					&				&	$H$	&	$53$		&	$0.52$	&	$24.18$ ($23.74$)	\\
CF2				&	16:02:06.80		&	-17:25:31.70	&	$Y$	&	$122$	&	$0.53$	&	$25.78$ ($25.34$)	\\
				&					&				&	$H$	&	$53$		&	$0.64$	&	$24.90$ ($24.46$)	\\
				&					&				&	$Ks$	&	$33$		&	$0.52$	&	$24.35$ ($23.91$)	\\ 
\hline
\end{tabular}
\label{obs}
\begin{list}{}{}
\item[$^{\mathrm{a}}$] Coordinates are given at the center of the HAWK-I $Y$-band image.
\end{list}
\end{table*}

The two HzRG fields were imaged between April 2008 and August 2009 in Service Mode with VLT/HAWK-I. HAWK-I
is a wide-field imager on UT4 with a field of view of 
$7.5\arcmin \times 7.5\arcmin$ equipped with a mosaic of four Hawaii 2RG $2048 \times 2048$ pixel 
detectors separated by a gap of $15\arcsec$. The pixel scale is $0.1064\arcsec$. 
The field of MRC~1017-220 was observed in Spring 2008
for $122$~min in the $Y$ band ($\lambda_C = 10210$\AA), $53$~min in the $H$ band 
($\lambda_C = 16200$\AA) and $33$~min in the $Ks$ band ($\lambda_C = 21460$\AA). 
MRC~0156-252 was also observed in the same set of filters:
$200$~min in the $Y$ band (Autumn 2008), $47$~min in the $H$ band and $33$~min in the $Ks$ band (August 2009).
In order to avoid the gaps between the chips and to have a deeper 
coverage of the immediate surroundings of the HzRGs, MRC~1017-220 and MRC~0156-252 
were placed near the center of one of the chips. 

Two control fields (hereafter CF1 and CF2, respectively centered on R.A.: 11:39:59.66, 
Dec.: -11:24:29.5 and R.A.: 16:02:06.80, Dec.: -17:25:31.7) were also observed from 
May to September 2008 in the $Y$, $H$ and $Ks$ bands. 

Part of each $Y$-band observation was observed in photometric sky conditions. 
A standard star, selected from the United Kingdom Infrared Telescope (UKIRT) 
faint standards list \citep{Hawarden2001}, was observed immediately after the 
science data to flux calibrate the $Y$-band image.

Between January and March 2008, the entrance window of HAWK-I suffered a degradation
and, as a consequence, the shadow of the camera spider became visible on the data. The data
thus contain the convolution of the spider shadow with the rotating pupil image, resulting in a 
cross pattern which repeats the spider symmetry in the background\footnote[1]{See http://www.eso.org/observing/dfo/quality/HAWKI/Problems/ PupilGhosts.html and http://www.eso.org/sci/facilities/paranal/ instruments/hawki/doc/HAWKI-NEWS-2008-07-11.pdf for details on the HAWK-I entrance window problem}. 
As advised by the HAWK-I User Support Team, we reduced the time for each sub-integration to attenuate
the cross pattern. However, some HAWK-I data taken between April and August 2008 show 
an increase in the sky noise independently of the configuration of the observations and the 
cross pattern is therefore very hard to subtract (e.g.,~$H$ and $Ks$-bands of CF1).
During the data reduction phase, we optimized the subtraction of the cross pattern 
when removing the background. Unfortunately, the cross pattern in the $Ks$-band of CF1 
could not be properly subtracted and the image is unusable.

The data were reduced using the ESO/MVM (or `alambic') reduction 
pipeline\footnote[2]{http://archive.eso.org/cms/eso-data/data-packages/eso-mvm-software-package}
\citep{Vandame2004}.
The image processing followed standard near-infrared reduction steps : dark subtraction, division by
normalized sky flats, subtraction of the background, fringing correction and harmonization of the four 
chip gains. A distortion correction was applied to each chip using stars from the USNO-B1.0 
\citep{Monet2003} catalogue. The images were finally stacked using the same astrometric catalogue 
\citep[for full details on the pipeline, see][]{Vandame2004}. 

\subsubsection{Photometric calibration}

\begin{figure}[!t] 
\begin{center} 
\includegraphics[width=8cm,angle=0]{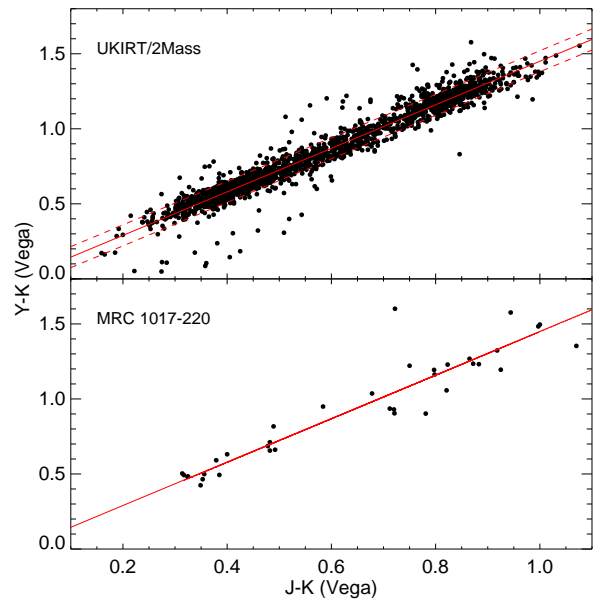}
\end{center}
\caption{Colour-colour diagram ($Y-K$ vs $J-K$) for the stars from a combined 
UKIDSS/2MASS catalogue (top panel) and for the 2MASS stars of the 
MRC~1017-220 field (bottom panel), shown as an example for the calibration 
of the $Y$-band HAWK-I data. Stars are well fitted by an
empirical colour-colour relation, $(Y-K)_{Vega} = 1.45\times(J-K)_{Vega}$ 
(solid line). The standard deviation ($0.07$) of the distribution is shown by the dotted lines.}
\label{calib}
\end{figure}

The $H$ and $Ks$-bands were flux calibrated using the 2MASS point source catalogue
\citep{Skrutskie2006} for objects with $11 \le K \le 14.5$ (i.e.~$22$ stars for MRC~1017-220, $13$
for MRC~0156-252, $9$ for CF1 and $37$ for CF2). 2MASS stars total magnitudes were 
estimated using SExtractor Kron aperture (MAG\_AUTO parameter). The derived zeropoints are 
accurate to $0.06$ magnitudes. The $2\sigma$ and $3\sigma$ detection limits of our images, determined in 
randomly distributed $1.5\arcsec$ diameter apertures, are reported in Table~\ref{obs}.

We derive the zeropoint of the $Y$-band using the standard stars observations and then 
refine it using empirically derived near-infrared colour relations for stars. 
Using the Wide Field Infrared Camera (WFCAM) Science Archive\footnote[3]{http://surveys.roe.ac.uk/wsa/} 
which holds images and catalogues of the UKIRT Infrared Deep Sky Surveys (UKIDSS), 
we retrieve stars from a $1.5\times1.5$ square degree region (centered around 
R.A.: 14:00:00, Dec.: 10:00:00) and match the $Y$-photometry of UKIDSS (data 
release 5) with 2MASS $J$ and $K$-photometry. We extract stars with accurate 
photometry in $Y$, $J$ and $Ks$ (less than $0.05$ magnitude errors; $1707$ stars). Fig.~\ref{calib} 
(top panel) shows the location of these stars in a $(Y-K)_{Vega}$ vs $(J-K)_{Vega}$ colour-colour 
diagram. Both UKIDSS and 2MASS use Vega photometric systems so Vega magnitudes 
are used in this analysis for consistency. The colour-colour relation for stars is 
well-fit by a simple linear function $(Y-K)_{Vega} = 1.45\times(J-K)_{Vega}$ (standard
deviation of $0.07$). We extract the 2MASS stars from the HzRGs fields 
and the two control fields 
and refine the zeropoints previously derived from the standard stars observations using the above 
colour-colour relation See the bottom panel of Fig.~\ref{calib} for an example of the calibration method 
used for the MRC~1017-220 field. The offsets applied to the initial zeropoint are less than $0.03$~mag. 
We estimate an average $0.06$~mag uncertainty in the 
$Y$-band photometry of MRC~1017-220, CF1 and CF2 and $0.07$~mag for MRC~0156-252.
This last field contains fewer stars and thus its photometry is slightly more uncertain. 
The $2\sigma$ and $3\sigma$ limiting magnitudes of the $Y$-band data (determined in random $1.5\arcsec$ 
diameter apertures) are given in Table~\ref{obs}.

\subsubsection{Archival GOODS-S data}

The Southern field of the Great Observatories Origins Deep Survey \citep[GOODS-S; ][]{Dickinson2003} 
was observed in $J$, $H$ and $Ks$ using VLT/ISAAC from October 1999 to January 2007. The data 
were reduced with the ESO/MVM pipeline by the GOODS team \citep{Retzlaff2010}. The final 
data release, available since September 2007, includes $24$ ISAAC fields in $H$ and $26$ ISAAC 
fields in $Ks$ as well as the final $H$ and $Ks$ combined mosaics\footnote[4]{The reduced single field images and 
final mosaics are publicly available at http://archive.eso.org/archive/adp/GOODS/ ISAAC\_imaging\_v2.0/goodsreq.html} 
covering respectively $159.6$ and $173.1$ arcmin$^2$. 

The GOODS-MUlticolor Southern Infrared Catalog\footnote[5]{Publicly available at http://lbc.mporzio.astro.it/goods/goods.php}
\citep[GOODS-MUSIC;][]{Grazian2006B, Santini2009} is a multiwavelength catalogue 
of GOODS-S, covering $143.2$ arcmin$^2$ and cross-correlated optical ($u$, $b$, $v$, $i$, 
$z$ from Hubble/ACS and VLT/VIMOS), near-infrared (see above), mid-infrared ({\it Spitzer}/Irac [$3.6$], 
[$4.5$], [$5.8$], [$8.0$] and {\it Spitzer}/MIPS $24\mu$m). Spectroscopic redshifts are available
for $12$\% of the sources. GOODS-MUSIC provides photometry for $12$ of the $24$ fields in $H$ and
$22$ of the $26$ fields in $Ks$. We make use of the GOODS-MUSIC photometry to calibrate the $H$ and $Ks$  
GOODS-S mosaics. The $2\sigma$ and $3\sigma$ limiting magnitudes in random 
$1.5\arcsec$ diameter aperture are $25.36$ ($24.92$) and $25.20$ ($24.76$) for $H$ and $Ks$.

\begin{figure}[!t] 
\begin{center} 
\includegraphics[width=9.5cm,angle=0]{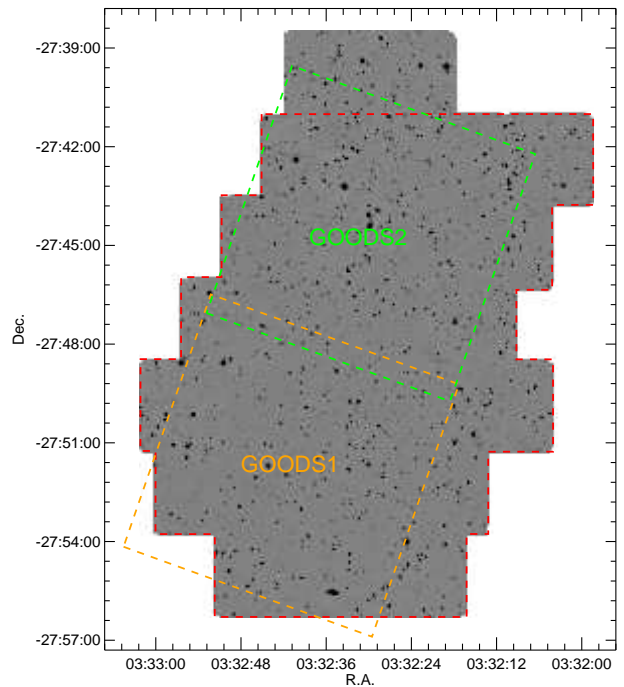}
\end{center}
\caption{$Ks$-band combined mosaic of the $26$ tiles observed by VLT/ISAAC in GOODS-S.
The $24$ fields also observed in $H$ are highlighted by the red dashed line. The two VLT/HAWK-I
$Y$-band observed as part of the science verification of the instrument are overlaid (orange: GOODS1,
green: GOODS2).}
\label{goods}
\end{figure}

As part of the VLT/HAWK-I Science Verification programs, two sub-fields of GOODS-S were 
observed in $Y$ in December 2007, centered respectively on R.A.: 03:32:40.92, Dec.: -27:51:41.6
($355$~min; PI: Fontana, A., GOODS1, hereafter) and R.A.: 03:32:29.71, Dec.: -27:44:38.6 
($145$~min; PI: Venemans, B., GOODS2, hereafter). See also \citet{Castellano2010} for further 
details on the data. The seeing of these images is consistent with the one of the ISAAC $H$ and 
$Ks$ images i.e.,~$\sim0.5-0.6\arcsec$. Fig.~\ref{goods} shows the $Ks$-band mosaic of GOODS-S 
with the $H$-band (red) and the two $Y$-band fields, GOODS1 (orange) and GOODS2 (green) overlaid. 
We reduce the data using ESO/MVM pipeline. The images are astrometrically calibrated using a 
source catalogue extracted from the GOODS $Ks$-band mosaic. To flux calibrate the images, we 
select all objects in the field classified as stellar in the NASA/IPAC Extragalactic Database (NED).  
Broad band photometry for these stars are taken from the Multiwavelength Survey by Yale-Chile 
(MUSYC) catalogues \citep{Gawiser2006}. The broadband SED of the stars are fit using stellar  
templates from the BPGS spectrophotometric atlas \citep{Hewett2006}. For each star, the best fitting 
SED provides the $Y$ magnitude of the star and, combined with the flux measured in the image, a  
zeropoint. The image zeropoint is derived from the average of the individual zeropoints and has an 
uncertainty of 0.05 magnitudes. We also independently determine the $Y$-band zeropoint using 2MASS 
stars in the GOODS-S field ($8$ stars) and the relation $Y-K=1.45(J-K)$ used for photometric calibration in \S2.2.2
and find results consistent within $0.03$ mag.
The $Y$-band $2\sigma$ ($3\sigma$) limiting magnitudes in random $1.5\arcsec$ diameter aperture are 
$26.94$ ($26.50$) and $26.41$ ($25.97$) for GOODS1 and GOODS2 respectively.

\section{Source extraction}

HAWK-I data, especially those taken prior to May 2009, contain crosstalk between 
the amplifiers of the chips i.e., all the sources are repeated on some of the other 
$32$ amplifiers producing a series of crater-like artifacts arrayed horizontally
\citep{Finger2008}. Due to its intensity, only crosstalk produced by the brightest 
objects is observed out of the background, e.g., in the image, all sources with 
$K_{2MASS,Vega}<15.5$. Regions affected by crosstalk were masked before 
extracting the source catalogues. Crosstalk is more prominent in the $Y$-band 
since the image is the deepest and stars are brighter in bluer bands. We therefore 
first identify by eye the crosstalk in the $Y$-band.
A map is created to flag the crosstalk-affected pixels by $4\arcsec \times 
4\arcsec$ squared masks. The flag area accounts for less than $2$\% of the 
final $Y$ images for MRC~1017-220 and CF1 and less than $1$\% for GOODS-S. 
The CF2 field contains numerous bright stars and $\sim4$\% of the final mosaic is flagged. 
We also flag regions affected by bright stars that dominate their surroundings as well as the 
noisy edges of the images.

The source detection was performed using SExtractor (Bertin \& Arnouts 1996)
\nocite{Bertin1996} with a detection threshold of $2\sigma$ independently for 
each filter. 
We used aperture magnitudes (SExtractor MAG\_APER) within a fixed $2.5\arcsec$ 
diameter aperture to measure colours. Based on the profile of 
stars in the $Ks$-band data of the HzRGs fields, we estimate that an aperture 
of $2.5\arcsec$ diameter contains about $95$\% of the source flux. Using 
such an aperture is therefore a good compromise between including as much 
flux from the source as possible, but limiting background contamination 
for faint objects. However, the fraction of a source flux contained in a $2.5\arcsec$ 
aperture strongly depends on the seeing of the image and we smoothed the 
images of different bands to the same seeing to ensure accurate colour measurements.

The 3-band images of MRC~1017-220 and CF2 have significantly different seeings
from filter to filter. We therefore smoothed our images to the worst seeing. The $H$ and 
$Ks$-band images of the field of MRC~1017-220 were smoothed to the $0.8\arcsec$ seeing 
of the $Y$-band. Similarly, the $Y$ and $Ks$-band images of CF2 were smoothed to the $0.64\arcsec$
seeing of the $H$-band. We detected the sources on the unsmoothed images and determined 
the aperture photometry on the smoothed images. We explain in \S5.3 how we handle upper limits 
in colours. Total magnitudes are determined using SExtractor parameter MAG\_AUTO on the 
original unsmoothed images.
 
All magnitudes were corrected for 
Galactic extinction (calculated for HAWK-I filters) using the dust maps of 
\citet{Schlegel1998} and assuming $R_V = A_V/E(B - V) = 3.1$ extinction law of 
\citet{Cardelli1989}. All the fields are at high galactic latitude ($b>20$). Corrections  
are small for the MRC~1017-220 field ($0.054$, $0.029$, $0.019$~mag in $Y$, $H$ and 
$Ks$ respectively), the MRC~0156-252 field ($0.013$, $0.007$, $0.004$~mag) 
and CF1 ($0.033$, $0.018$, $0.011$~mag). Due to a lot of dust along the line of 
sight, corrections for CF2 are, on the contrary, rather big 
($0.344$, $0.187$, $0.118$). We do not apply the negligible ($<0.01$ in $Y$ and 
$<0.005$ for $H$ and $Ks$) extinction corrections for the GOODS-S field.

\begin{figure}[!t] 
\begin{center} 
\includegraphics[width=6.2cm,angle=90]{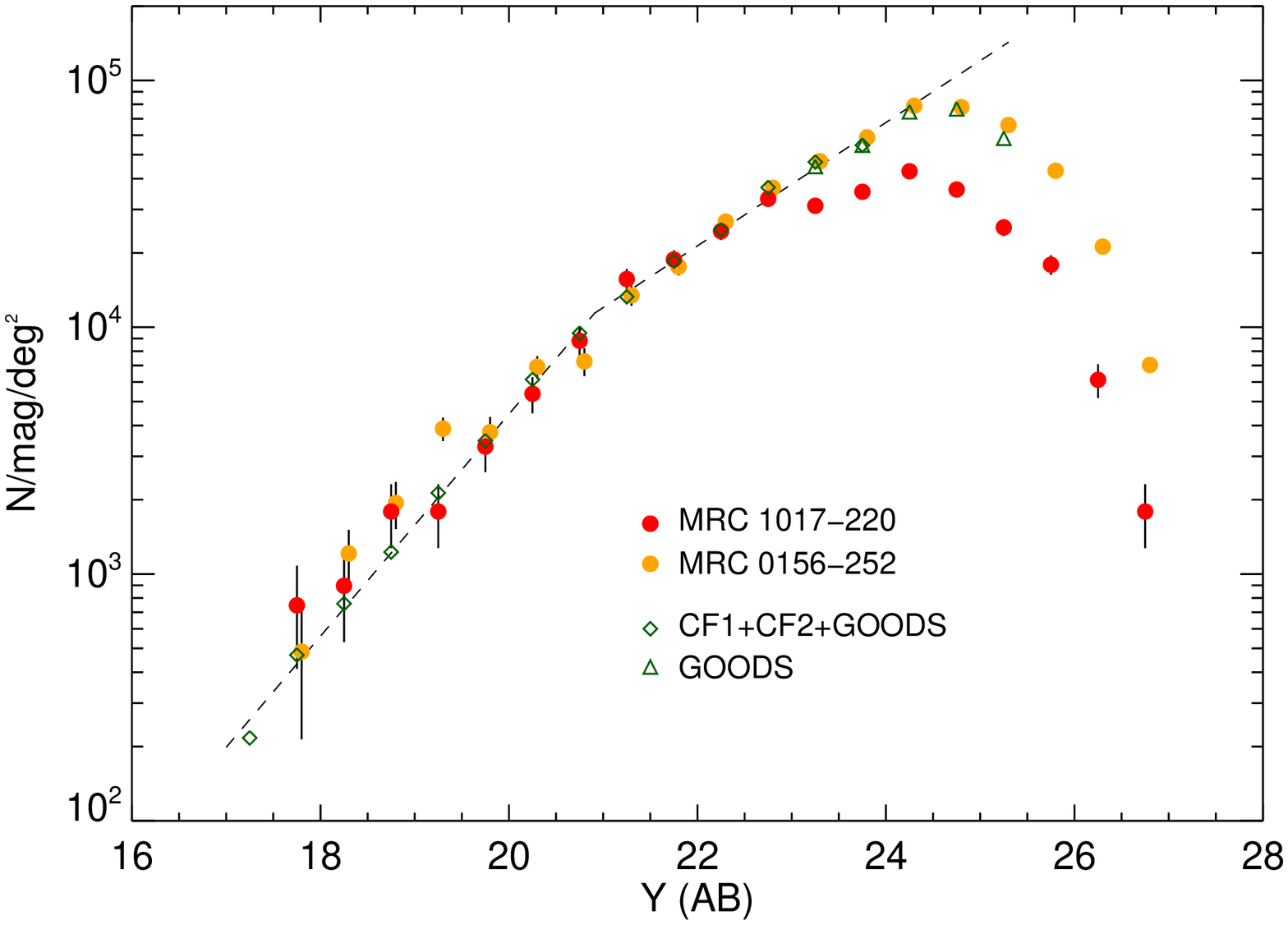}
\includegraphics[width=6.2cm,angle=90]{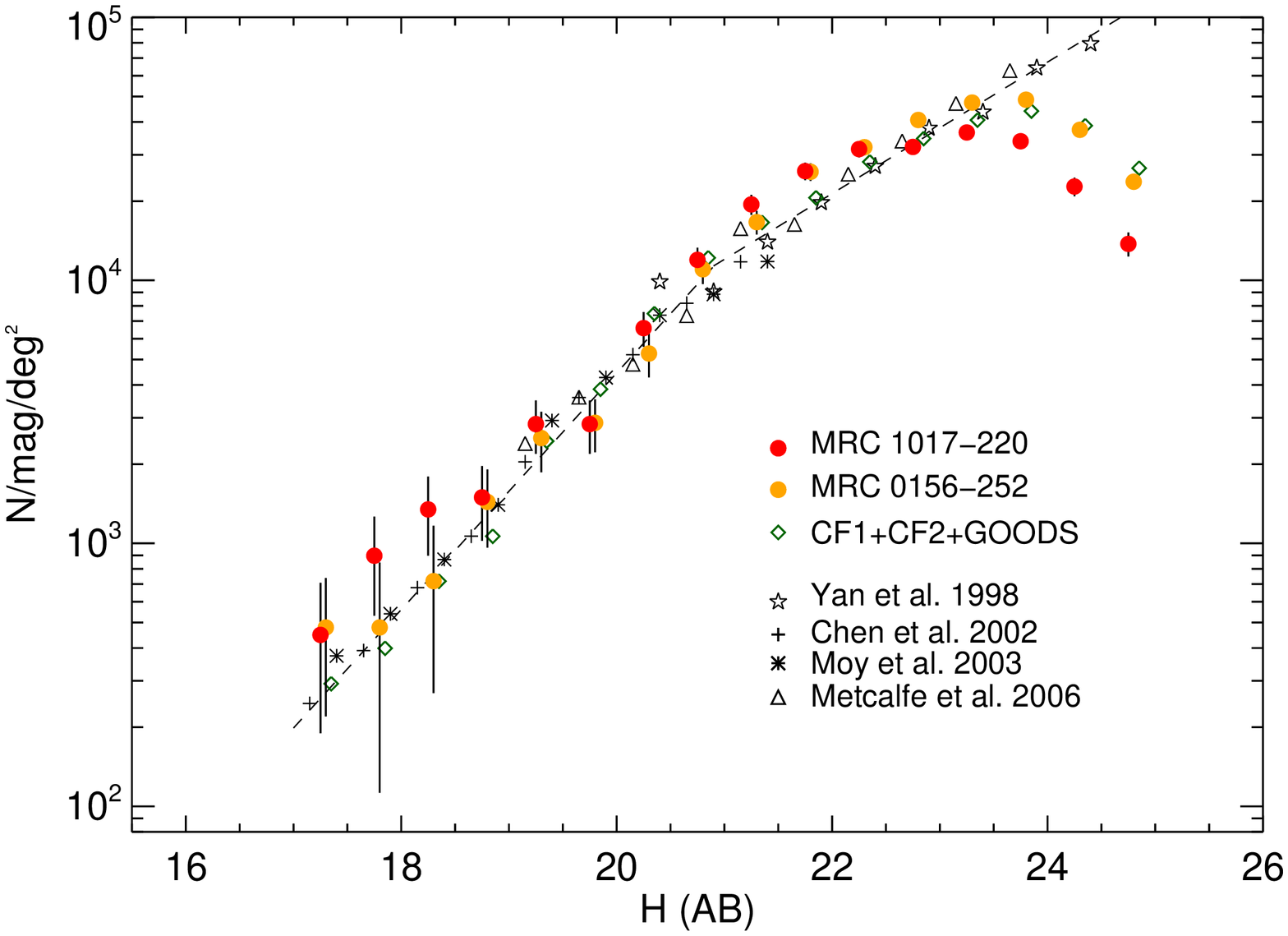}
\includegraphics[width=6.2cm,angle=90]{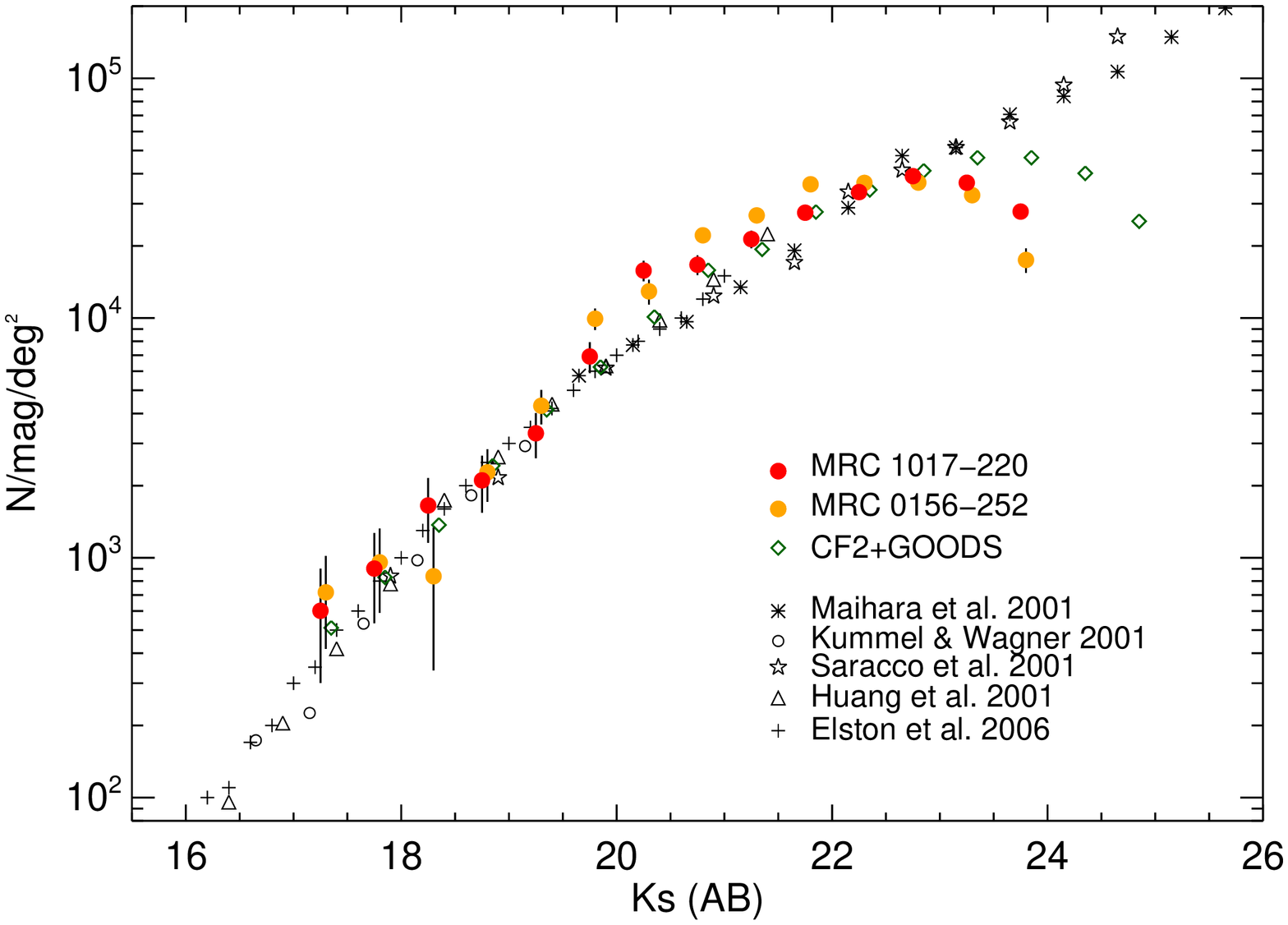}
\end{center}
\caption{Galaxy number counts in $Y$, $H$ and $Ks$ for the field around MRC~1017-220 (red dots),
MRC~0156-252 (orange dots) and the combined control fields (green diamonds). 
We do not apply completeness correction. We isolate galaxies from stars using 
the stellar index CLASS\_STAR in SExtractor. Counts from the literature are 
overplotted for $H$ and $Ks$. For the $Y$-band, we derive counts from the two 
GOODS-S $Y$ field and the two control fields (CF1 and CF2) until $Y<24$. 
Galaxy counts are also derived from GOODS-S only for sources with $23<Y<25.5$. 
The best  fit of the $Y$-band number counts by a two power-law model is shown 
by the dashed line (see \S4, also overplotted on the $H$-band 
number counts for comparison).}
\label{counts}
\end{figure}

We evaluated the completeness limits of the images using a IRAF \texttt{gallist} 
and \texttt{mkobjects} routines (\texttt{artdata} package) to simulate artificial 
galaxies, both elliptical 
and spiral galaxies. 
We chose a uniform distribution of galaxy morphologies with a minimum 
galaxy axial ratio \textit{b/a} of $0.8$ and a maximum half flux radius of $1$\farcs$0$. 
We adopted a de Vaucouleurs and an exponential disk surface brightness law for
ellipticals and spirals respectively. For both types of galaxies, we generated
catalogues of $5000$ objects and added them to the $Y$, $H$ and $Ks$ images, 
including Poisson noise. 
We determined how many artificial sources were
recovered using the same SExtractor configuration files used to detect the real sources. 
The $90$\% completeness limits for elliptical (spiral) galaxies are $24.0$ ($23.3$), 
$22.9$ ($22.2$) and $22.4$ ($21.7$) in $Y$, $H$ and $Ks$ respectively for MRC~1017-220, 
$24.4$ ($23.8$), $23.2$ ($22.5$) and $22.4$ ($21.8$) for MRC~0156-252.

\section{Galaxy number counts in $Y$, $H$ and $Ks$}

\begin{table}
\caption{Galaxy number counts in $Y$-band (N/(mag.deg$^2$))}
\label{countsY}
\centering
\begin{tabular}{c c c c c}
Mag. & 1017 & 0156 & CFs+GOODS-S & GOODS-S \\
(1) & (2) & (3) & (4) & (5) \\
\hline
17.75 & 747 &	485 & 470 & - \\
18.25 & 896 & 1213 & 759 & - \\
18.75 & 1792 & 1940 & 1228 & - \\
19.25 & 1792 & 3881 & 2131 & - \\ 
19.75 & 3286 & 3759 & 3468 & - \\
20.25 & 5377 & 6913 & 6141 & - \\
20.75 & 8812 & 7276 & 9464 & - \\
21.25 & 15682 & 13461 & 13293 & - \\
21.75 & 18818 & 17585 & 18531 & - \\
22.25 & 24494 & 26801 & 24816 & - \\
22.75 & 33156 & 36746 & 36845 & - \\
23.25 & 31066 & 47054 & 46742 & 44807 \\
23.75 & 35397 & 58817 & 54652 & 54695 \\
24.25 & - & 78948 & - & 74609 \\
24.75 & - & - & - & 76683 \\
\hline
\end{tabular}
\end{table}

We derive the differential galaxy number counts in the HzRG fields for each band 
(see Fig.~\ref{counts}). We only consider the deepest regions of the final images, i.e., 
we discard the shallowest regions resulting from the gap between chips (central `cross')
as well as the edges of the image i.e.~$12-13$\% of the images. 
We also derive the galaxy number counts combining the two control fields: CF1, CF2 
and the GOODS-S data for the $Y$ and $H$ bands. For the $Ks$ band, number 
counts are determined from CF2 and GOODS-S, since CF1 was not observed in $Ks$.
We separate galaxies from stars using the SExtractor CLASS\_STAR parameter. We test
the optimal values for this parameter for each image with stars from 2MASS and USNO
in flux bins of $1$~mag ($0.6<$CLASS\_STAR$<0.9$).

Table~\ref{countsY} reports (for the first time) the $Y$-band galaxy number
counts. Counts for the two HzRGs fields (columns 2 and 3) are given 
to the $90$\% completeness limit. 
Due to the lower completeness limits of CF1 and CF2, we first derive counts for 
$Y<24$ from the four fields (CF1+CF2+GOODS1+GOODS2) i.e., a total area of 
about $200$~arcmin$^2$ (see Table~\ref{countsY}, column 4). We also derive
the counts for $23<Y<25.5$ from GOODS-S only (column 5). 

For the two other bands, we compare the galaxy counts with results 
from the literature: \citet{Yan1998, Chen2002, 
Moy2003, Metcalfe2006} for $H$ and \citet{Maihara2001, Kummel2001, 
Saracco2001, Huang2001, Elston2006} for $Ks$. No attempt is made to correct
for the incompleteness, nor correct for differences of filter passbands (e.g.,~$K$ or $Ks$). 

Number counts have been frequently used to test models of galaxy evolution. Although such an analysis
is beyond the scope of this paper, we note however that the slope shown by the $Y$-band 
counts is consistent with the $H$-band. \citet{Imai2007} fitted their $J$-band 
counts (derived from the $AKARI$ North Ecliptic Pole survey) by two power-laws 
of the form $N(mag)=a\times10^{b(mag-15)}$ with
a break in the slope at $J_{Vega}\sim19.5$. Adopting a similar fitting function, we
find that the $Y$-band counts are well fitted by two power-laws: $a=25\pm5$; $b=0.45\pm0.02$ for
$Y<21$ and $a=380\pm5$; $b=0.25\pm0.02$ for $Y>21$. These best-fit slopes are plotted in Fig.\ref{counts} 
(top panel) and overplotted on the $H$-band counts (middle panel) for comparison. The $Y$ and $H$-band
counts are found to be consistent. Similarly to \citet{Imai2007}, we note that the $Ks$-band number counts show
a less abrupt change of slope at $19<K<21$ and cannot easily be modeled by two power-laws.

As shown in Fig.~\ref{counts} (top panel), the $Y$ number counts of the HzRG fields are in good agreement
with counts derived from the control fields. We note that there is a deficit of sources in the MRC~1017-220 
field at $Y>23$, which cannot be explained by a lack of depth of the data since 
the 90\% completeness limit is reached at $Y=24$~mag.
Interestingly, the galaxy counts of the targeted HzRG fields show an excess 
of galaxies in $H$ for $20.5<H<22.5$ where we find $27.1\pm0.6$\% ($22.3\pm0.6$\%) more sources 
in the field of MRC~1017-220 (MRC~0156-252) than our control fields (CF1+CF2+GOODS).
Similar results are found in the $Ks$-band number counts where an excess of sources with 
$19.5<Ks<22.5$ ($27.4\pm0.4$\%) is found in the field of MRC~0156-252 compared to our control fields 
(CF2+GOODS). An excess of sources with $19.5<Ks<21.5$ ($17.8\pm0.6$\%) is also seen in the field of 
MRC~1017-220. We do not observe such excesses in the $Y$-band (which is below the $4000$\AA~break
at $z\sim2$). This suggests that both HzRG fields contain an excess of red galaxies.

\begin{figure}[!t] 
\begin{center} 
 \includegraphics[width=8.2cm,angle=0]{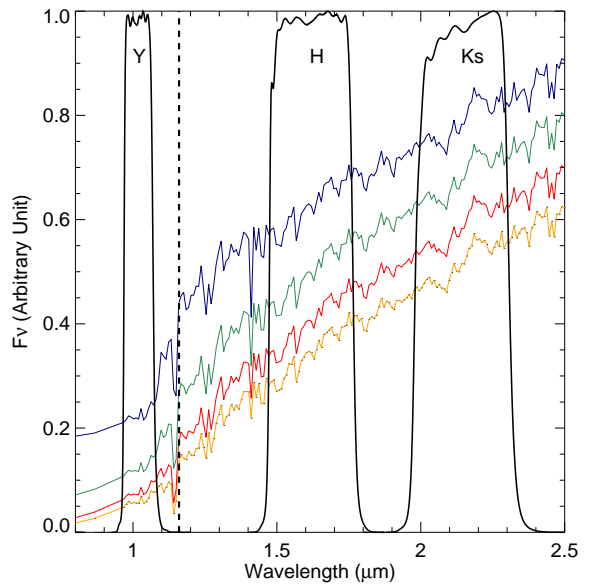}
\end{center}
\caption{Redshifted (to $z=1.9$) spectral energy distribution of models
of $2$~Gyr-old galaxies assuming an exponential declining star 
formation history with $\tau=0.1$, $0.3$, $0.5$ and $1$ (from yellow to blue; see
Fig.~\ref{BC} for colours). The transmission curves of the HAWK-I filters 
$Y$, $H$ and $Ks$ and the position of the restframe
$4000$\AA~break (dashed line) are overlaid.}
\label{templates}
\end{figure}

\section{Search for candidate cluster members at $z\sim2$}

\subsection{ $YHK$ colour selection of galaxies at $z>1.6$}

Colour criteria efficiently select high-redshift galaxies in a relatively narrow redshift 
range and permit to isolate potential cluster members associated with HzRGs.
At high redshifts ($2\lesssim z \lesssim3$), the position of the $4000$\AA~break 
in galaxies falls between the $J$
and $H$-band. Pure near-infrared colours have thus been used to isolate candidate
protocluster members at this redshift range, e.g.~$(J-K)_{Vega} > 2.3$,
designed by the FIRES team \citep{Franx2003} which allows them to select
Distant Red Galaxies (DRGs). However, these single colour cuts mostly pick out red 
passively evolving or dusty galaxies and miss galaxies with continuous star-formation 
with little or no dust. 

\citet{Kajisawa2006} defined a two-colour selection 
technique that combined $J$, $H$ and $K$ colours to select both red, passive and blue, star-forming 
galaxies at $z > 2$. This criterion ($J-K > 2\times(H-K)+0.5$ \& $J-K>1.5$; Vega system) is
almost insensitive to dust extinction since the reddening vector ($E(B-V)$) is parallel
to the colour selection. 

\begin{figure}[!t] 
\begin{center} 
\includegraphics[width=8.6cm,angle=0]{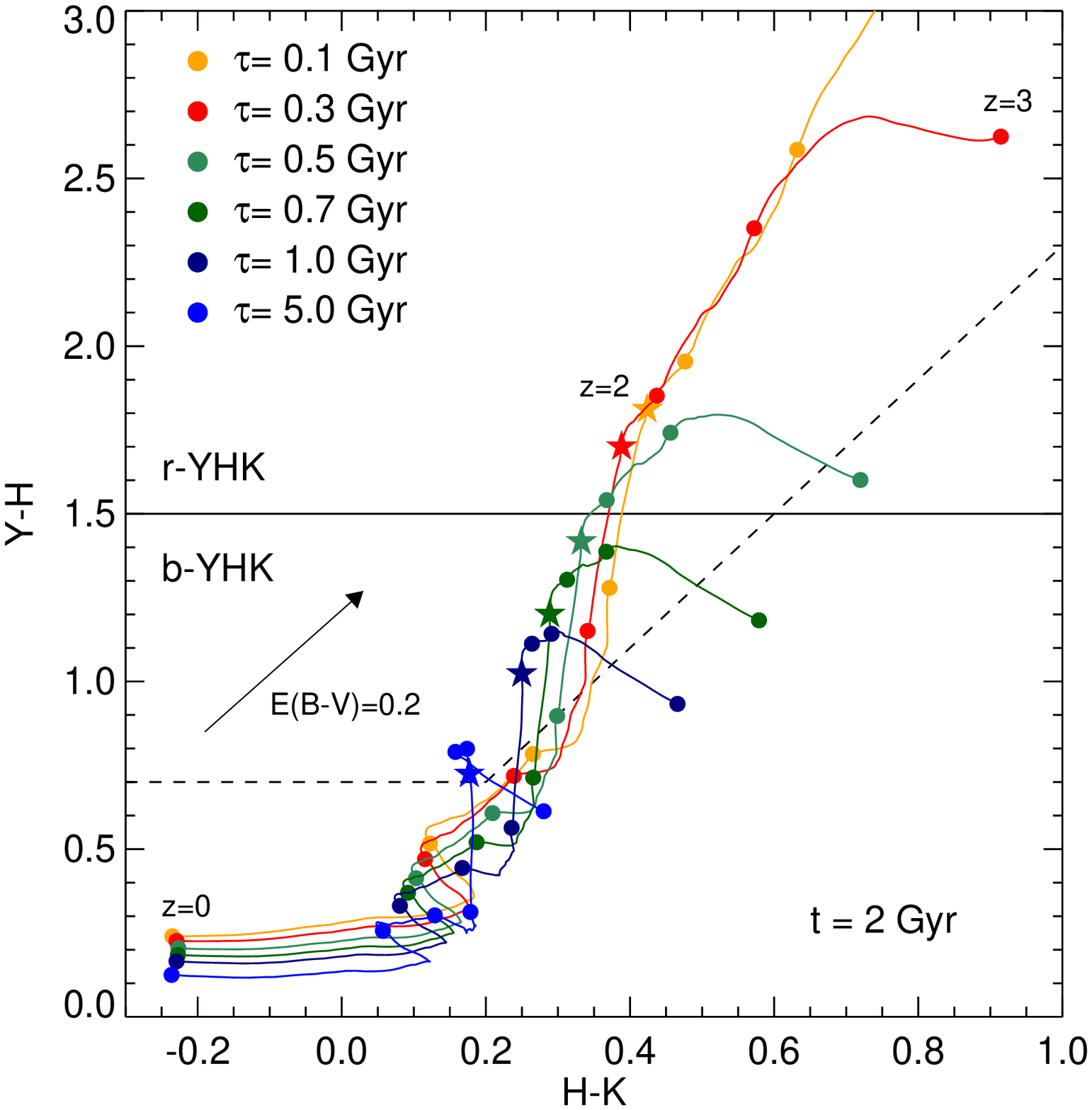}
\includegraphics[width=8.2cm,angle=0,bb=-2 14 564 350]{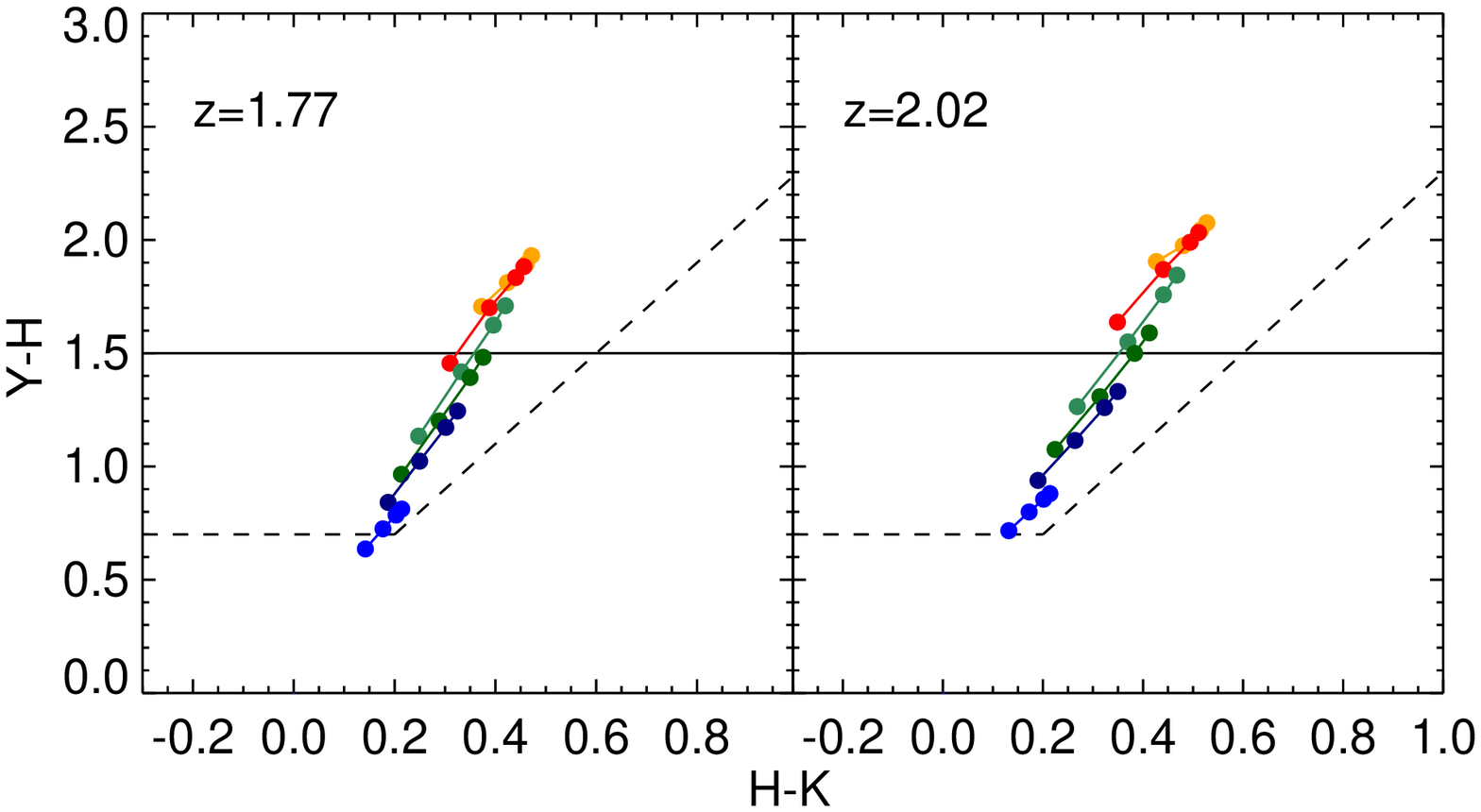}
\end{center}
\caption{\citet{Bruzual2003} model predictions of different stellar populations in a 
$Y-H$ vs $H-K$ colour-colour diagram (see text for details on the models). 
The dashed line accounts for the two-colour selection designed to isolate galaxies at $z>1.6$. 
The horizontal line ($Y-H>1.5$) shows the single-colour selection to separate red passively evolving 
galaxies (r-$YHK$ galaxies) from star-forming ones (b-$YHK$ galaxies). We assume 
an exponential declining star formation history with $\tau=0.1$, $0.3$, $0.5$, $0.7$, 
$1$ and $5$~Gyr (see legend).
{\it Top:} for increasing redshifts ($z=0$ to $z=3$, each $0.5$ bin marked by coloured points) and a 
constant age of $2$~Gyr, the stars corresponding to $z=1.77$. The black 
arrow indicates the reddening vector $E(B-V)=0.2$ as parameterized by \citet{Cardelli1989}.
{\it Bottom:} for various population ages ($t = 1.5$, $2$, $2.5$, $2.75$~Gyr, coloured 
points with $Y-H$ getting redder with ages) at $z=1.77$ and $z=2.02$.}
\label{BC}
\end{figure}

At the redshift of the targeted HzRGs, the $4000$\AA~break is between the $Y$ and the 
$J$-band. One can therefore use the $Y$-band instead of the $J$-band in a similar 
method to \citet{Kajisawa2006} to search for cluster members around MRC~1017-220 
and MRC~0156-252. We use \citet{Bruzual2003} models to determine the colours of 
different stellar populations at $z=1.77$ and $z=2.02$. In the models, we assume a solar 
metallicity and a \citet{Chabrier2003} initial mass function. Fig.~\ref{templates} shows dust-free 
SED predictions for galaxies with an exponentially declining star formation history with 
$\tau=0.1$ to $\tau=1$~Gyr at $z=1.9$. The HAWK-I transmission filters (black curves) 
and the position of the $4000$\AA~break at $z=1.9$ (dashed line) are also overlaid. 
We note that the $Y$ and $H$ filters bracket \textit{restframe} $4000$\AA. The $Y$ filter is 
narrow (width=$0.1\mu$m, about $3$ times narrower than $H$ and $Ks$). The 
$Y-H$ colour is therefore very sensitive at selecting galaxies at the targeted redshifts. 

Fig.~\ref{BC} shows the same \citet{Bruzual2003} SED models in a $Y-H$ vs $H-K$ 
colour-colour diagram at various redshifts, with a constant population age of $2$~Gyr (top panel), 
or at various population ages ($t = 1.5$ to $2.75$~Gyr) at $z=1.77$ and $z=2.02$ (bottom panel). 
At $z=0$, galaxies have consistently low $Y-H$ colours ($0<Y-H<0.3$). Beyond $z\sim1.4$, 
the $Y-H$ colour becomes redder when the $4000$\AA~break enters the $Y$-band. 
The $H-K$ colour stays almost constant until $z\sim2.5$ when the $4000$\AA~break 
enters the $H$ band. By $z>3$, the $Y-H$ ($H-K$) colour becomes 
rapidly bluer (redder) than those of galaxies at $1.5<z<3$. This general trend is observed 
for all models. However, the variations on both the $Y-H$ and $H-K$ colours are 
more pronounced for galaxies with short declining star-formation histories i.e.~those with 
stronger $4000$\AA~breaks. Galaxies with $\tau<0.5$Gyr and $z>1.6$ have  
$Y-H>1.5$. Galaxies with longer star-formation histories never become so red.

\begin{figure}[!t] 
\begin{center} 
\includegraphics[width=9.7cm,angle=0,bb=40 20 650 580]{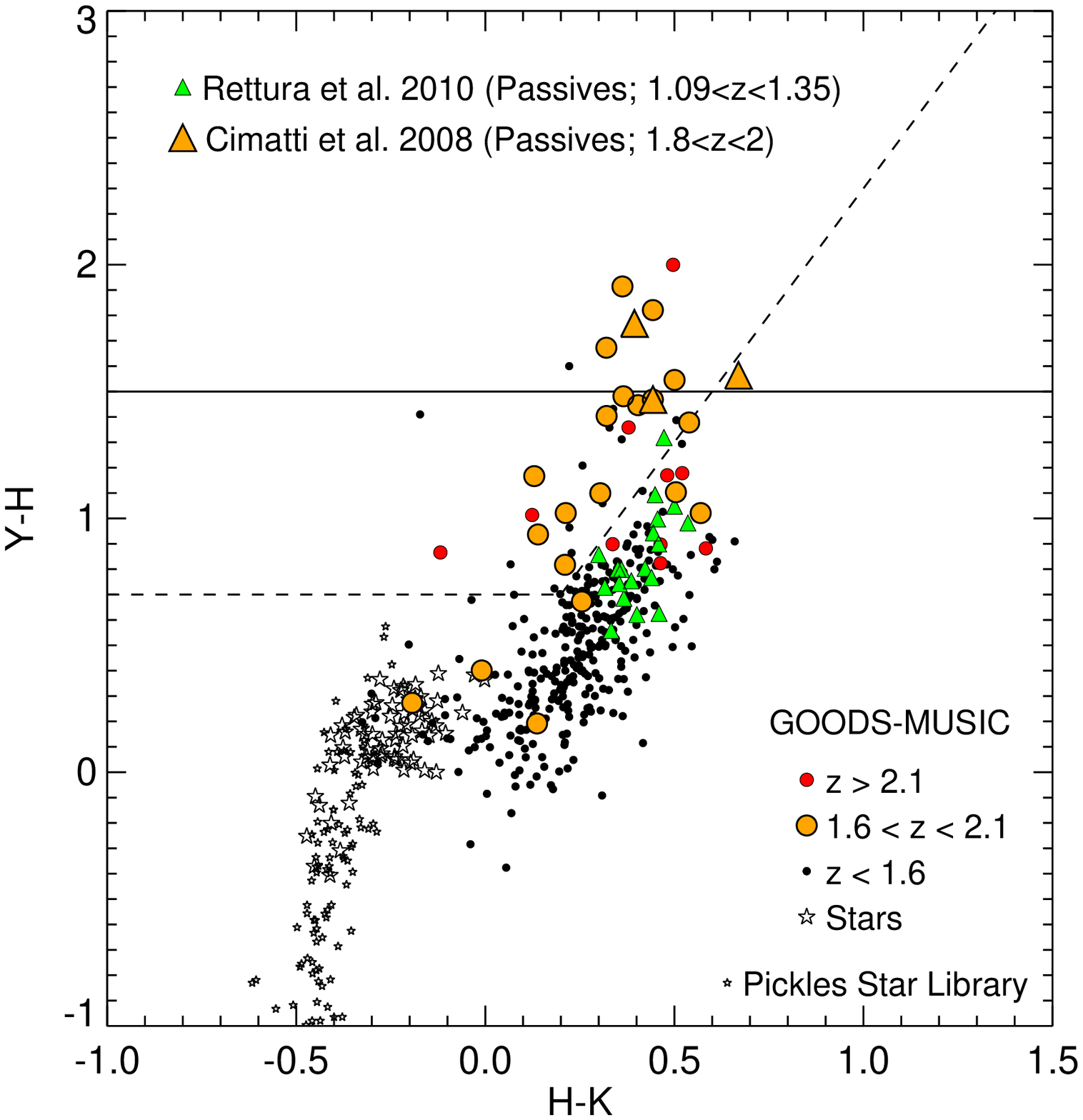}
\includegraphics[width=9.7cm,angle=0,bb=40 20 650 580]{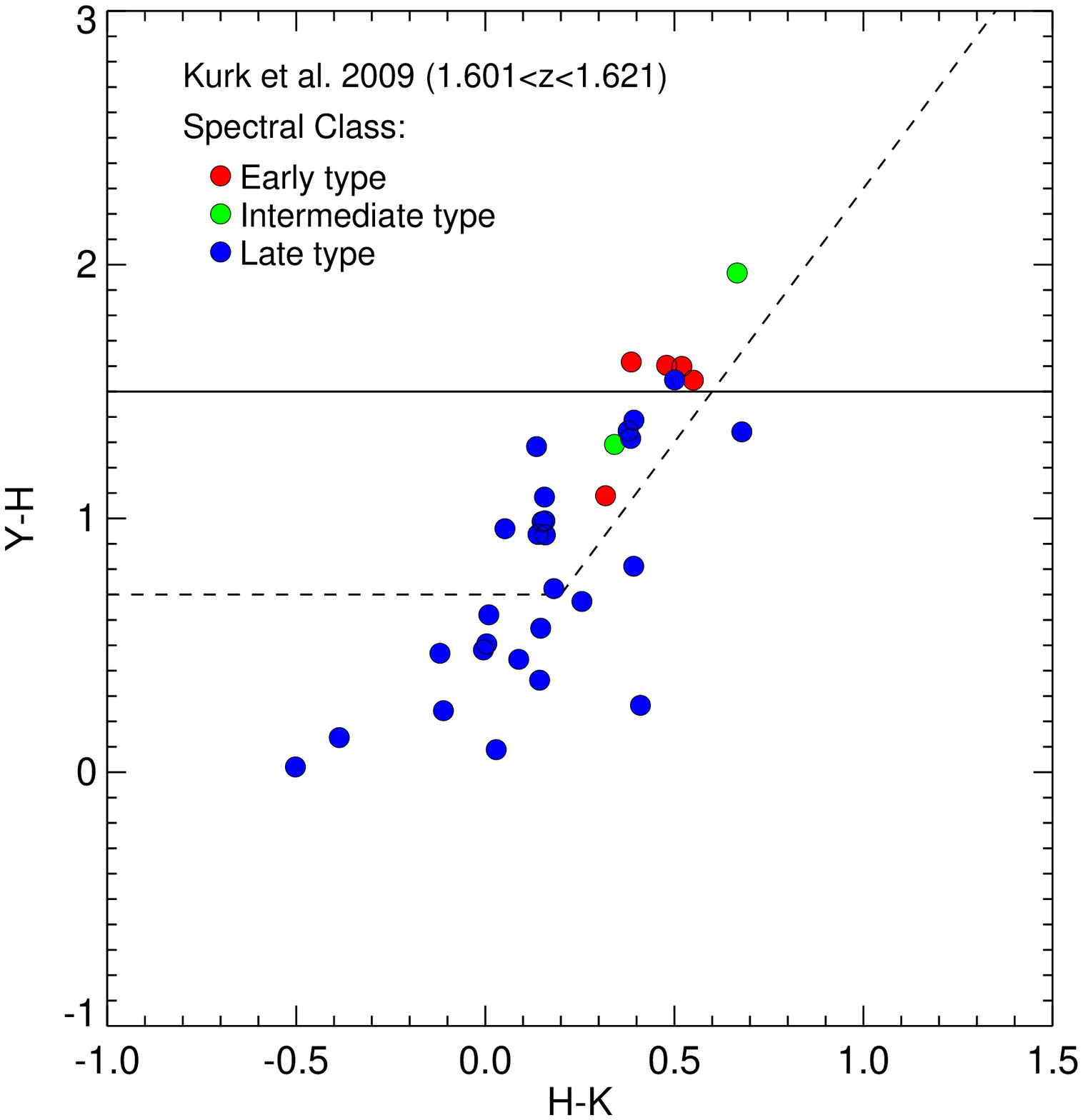}
\end{center}
\caption{$YHK$ colour-colour diagrams of sources with spectroscopic redshifts
from the literature (GOODS-S field). The selection criteria (1) and (2) are shown 
by the dashed and solid horizontal lines respectively. {\it Top:} Sources with spectroscopic 
redshifts from the GOODS-MUSIC catalogue (see legend for symbols) and 
stars from the Pickles~1998 stellar spectra flux library. We overplot early-type galaxies
from \citet{Rettura2010} at $z<1.5$ as well as the three passive galaxies at $z>1.8$ from 
\citet{Cimatti2008} (green and orange triangles respectively). {\it Bottom:}  Members of the overdensity at $z\sim1.6$ found in GOODS-S 
\citep{Kurk2009}. Colours of the symbols account for the spectral class
of the group members: red, green and blue for early-type, early-type with signs of
star formation and late type galaxies respectively.}
\label{GOODSzspec}
\end{figure}

We define a new colour criterion, analogous to the \citet{Kajisawa2006} two-colour 
$JHK$ selection technique, to select galaxies at $1.6 < z < 3$ (Fig.~\ref{BC}, dashed line):

\begin{equation}
Y-H \ge 0.7 \cap Y-H \ge 2 \times (H-K) +0.3
\end{equation}

A single-colour criterion is defined to separate red galaxies with old stellar population from blue
star-forming galaxies:

\begin{equation}
Y - H \ge 1.5
\end{equation}

similar to the single-colour criterion $(J-K)_{Vega}>2.3$ that selects DRGs at 
$z>2$. This criterion is shown in Fig.~\ref{BC} as the horizontal solid line. The cut at $Y-H=1.5$
was optimized to isolate the reddest population of galaxies at $z>1.6$. 
By analogy with \citet{Kodama2007} notations (r-$JHK$ and b-$JHK$) for
galaxies selected by the \citet{Kajisawa2006} near-infrared criteria, galaxies 
selected by equations (1)+(2)  are referred to as r-$YHK$ galaxies and galaxies selected by
equation (1) with $Y-H<1.5$ are referred to as b-$YHK$ galaxies.

The $YHK$ criterion is insensitive to dust extinction since it has been 
defined parallel to the reddening vector $E(B-V)$ (see Fig.~\ref{BC}; black 
arrow). However, the dusty star-forming galaxies at $z>1.6$ will have similar 
colours to non dusty, passively 
evolving galaxies, and will also be selected by the r-$YHK$ criterion.
Our criterion therefore does not enable us to clearly distinguish the red passive 
galaxies from the dusty star-forming ones.

\subsection{Reliability of the $YHK$ colour selection}

\begin{figure}[!t] 
\begin{center} 
\includegraphics[width=7.2cm]{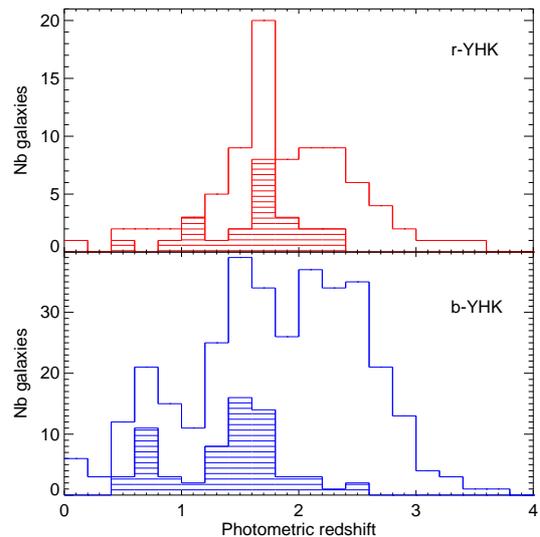}
\end{center}
\caption{Photometric redshift distribution of the $YHK$-selected sources 
for r-$YHK$ galaxies (top panel) and b-$YHK$ galaxies (bottom panel) in GOODS1+GOODS2 
for both the $2\sigma$ detection limits (solid histogram) and the limit of completeness
of the HAWK-I data (filled histogram). Photometric redshifts are from Santini et al.~2009.}
\label{zsel}
\end{figure}

We examine the $YHK$ colours of galaxies in GOODS-S using the spectroscopic redshifts available 
in the GOODS-MUSIC catalogue (see \S2.2.2) and combining the two archival $Y$-band (GOODS1 
and GOODS2) images with the $H$ and $Ks$ GOODS-S mosaics. We select sources that have a reliable 
photometry in $H$ and $Ks$ (less than $0.1$ magnitude errors), are detected ($2\sigma$) in the 
two $Y$-band images, and have a reliable spectroscopic redshift (flag 0: very good or 1: good).
Fig.~\ref{GOODSzspec} (top panel) shows the distribution of these sources in the $YHK$ 
colour-colour diagram i.e., $72$ stars (stars), $316$ galaxies at $z<1.6$ (black dots), 
$20$ galaxies with $1.6<z<2.1$ (orange circles) and $10$ galaxies with $z>2.1$ (red circles). 
AGN (defined as `BLAGN' or `NLAGN' in GOODS-MUSIC) are not considered here. 
We also overplot colours of stars from the digital stellar spectra library from \citet{Pickles1998}.
We note that $13/20$ ($65$\%) galaxies at $1.6<z<2.1$ are selected by the $YHK$ criteria. 
The criterion efficiently removes foreground galaxies i.e.,~only $14/316$ 
($<5$\%) galaxies with $z<1.6$ are found in the $YHK$-selected region.
$4/10$ ($40$\%) of galaxies with $z>2.1$ are selected by the $YHK$ criterion indicating
that the selection will be notably contaminated by background objects.

\citet{Rettura2010} present a sample of $27$ early-type galaxies with $1.09<z<1.35$ 
found in the GOODS-S field. $18$ of them are imaged and detected ($>2\sigma$) in the $Y$ 
band (see green triangles in Fig.~\ref{GOODSzspec}) and have colours consistent with models 
predictions. Only one of them (the highest redshift source at $z=1.35$) is selected by the $YHK$ 
criterion showing that contamination by lower redshift passively evolving galaxies is very small.

\citet{Cimatti2008} studied a sample of $13$ old, passive galaxies at $z>1.4$ found in the 
northern part of GOODS-S covered by the Galaxy Mass Assembly ultra-deep Spectroscopic 
Survey \citep[GMASS;][]{Kurk2008A}. Ten of these objects have $z>1.6$ including three at 
$z>1.8$. The seven others are part of an overdensity at $z\sim1.6$ presented in Kurk et al.~(2009). 
The three passive galaxies at $z>1.8$ are well detected in $Y$ (in the GOODS2 field), $H$ 
and $Ks$ (see orange triangles in Fig.~\ref{GOODSzspec}, top panel). Two of them have $Y-H>1.5$. 
The third one is a b-$YHK$ galaxy, but has a red $Y-H$ colour, close to the selection limit.

\citet{Kurk2009} discovered a galaxy overdensity at $z=1.6$, with $42$ spectroscopically 
confirmed members in the GMASS area. Five galaxies have an early-type spectral class,
two galaxies are at an intermediate stage (early-type but with sign of star formation; 
intermediate type, hereafter) and $35$ are late type galaxies. All $42$ members are 
detected in the GOODS2 $Y$-band. Eight late-type galaxies have magnitude
errors larger than $0.1$~mag in either $Y$, $H$ or $Ks$ and will not be considered in our analysis.
Fig.~\ref{GOODSzspec} (bottom panel) shows the colours of the $34$ remaining sources 
in the $YHK$ colour-colour diagram, the colours of the symbols indicating their spectral 
class. 

The five early-type galaxies and the two `intermediate type' galaxies are also reported in 
\citet{Cimatti2008} as passive galaxies. All seven sources are selected by the $YHK$ 
criterion. Four early-type and one `intermediate type' are r-$YHK$ galaxies and the last 
two sources are b-$YHK$ galaxies. $12/27$ ($44\%$) of the late type galaxies are 
selected by the criterion showing its limitation at $z\le1.6$. We also note that 
the [OII]$\lambda3727$\AA~doublet falls in the $Y$-band at $1.6<z<2$. If present, and has a 
large equivalent width, the emission line could bias the source $Y$-band magnitude and 
thus its $Y-H$ colour, with the galaxy appearing bluer than expected. Since the majority ($>80$\%) 
of the late-type galaxies from Kurk et al.~(2009) shows the [OII] line in their spectra, this could 
explain the blue colours of some of them.

\begin{figure*}[!t] 
\begin{center} 
\includegraphics[width=5.7cm,angle=0,bb=0 10 500 600]{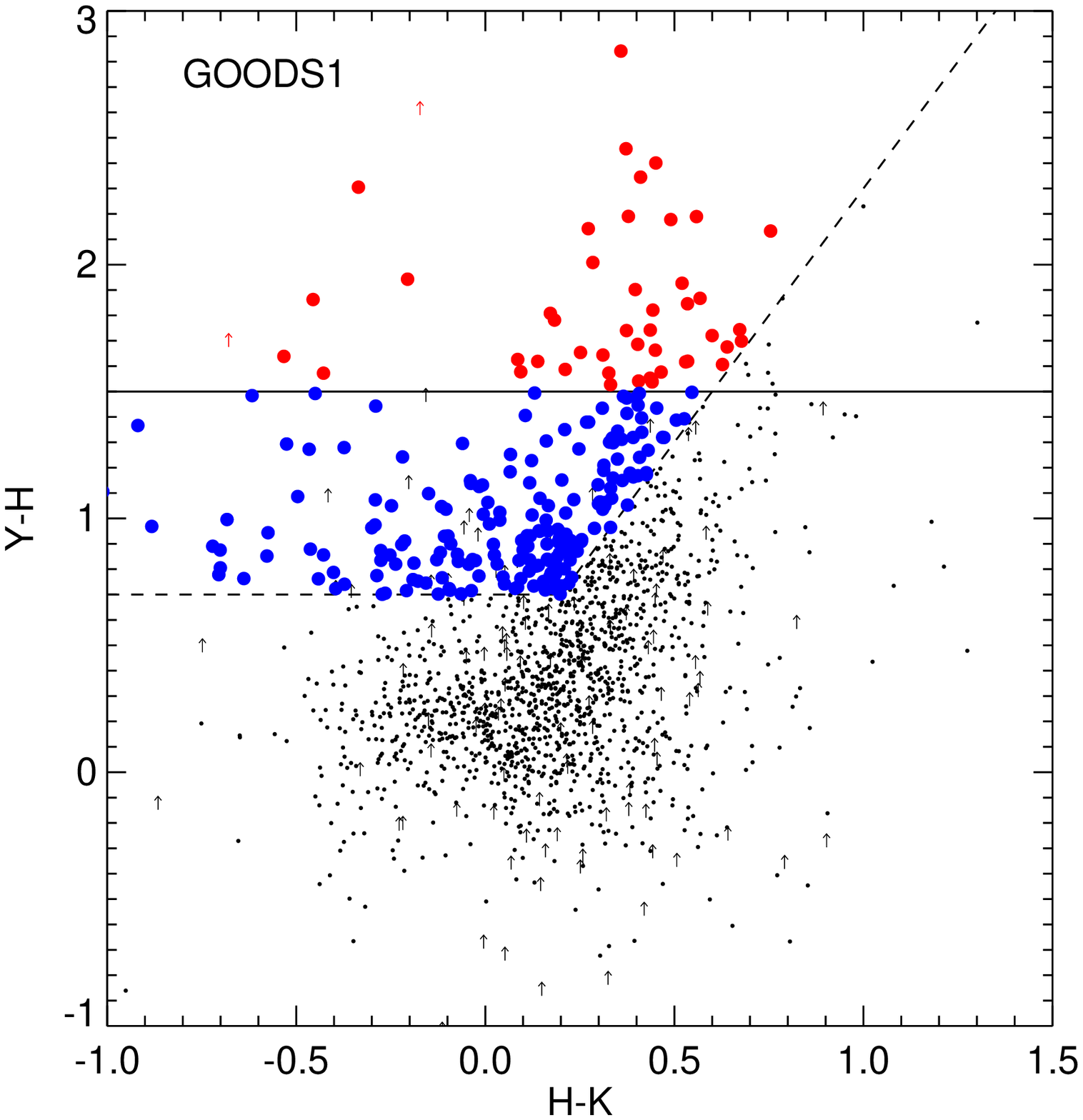}
\includegraphics[width=5.7cm,angle=0,bb=0 10 500 600]{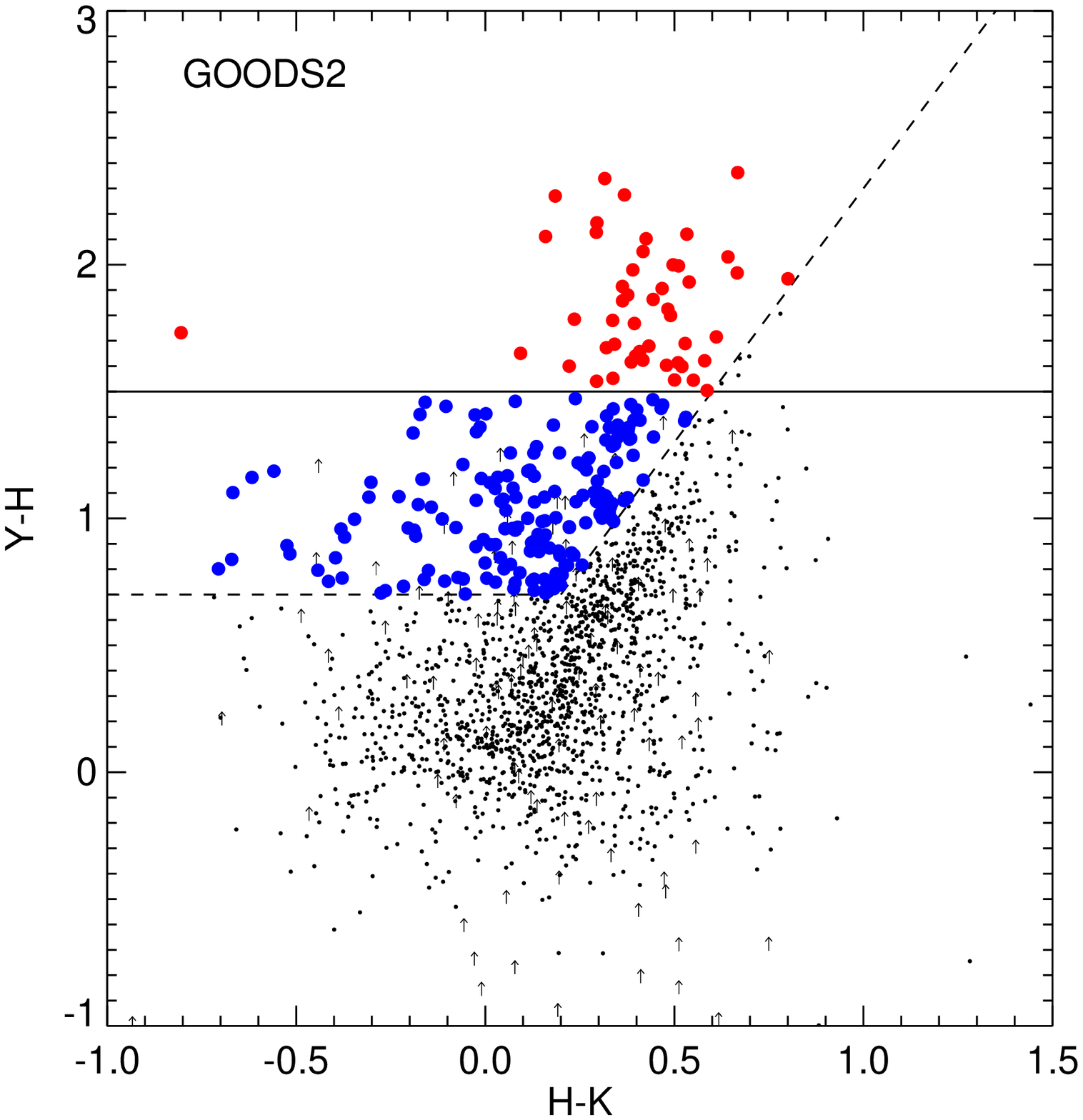}
\includegraphics[width=5.7cm,angle=0,bb=0 10 500 600]{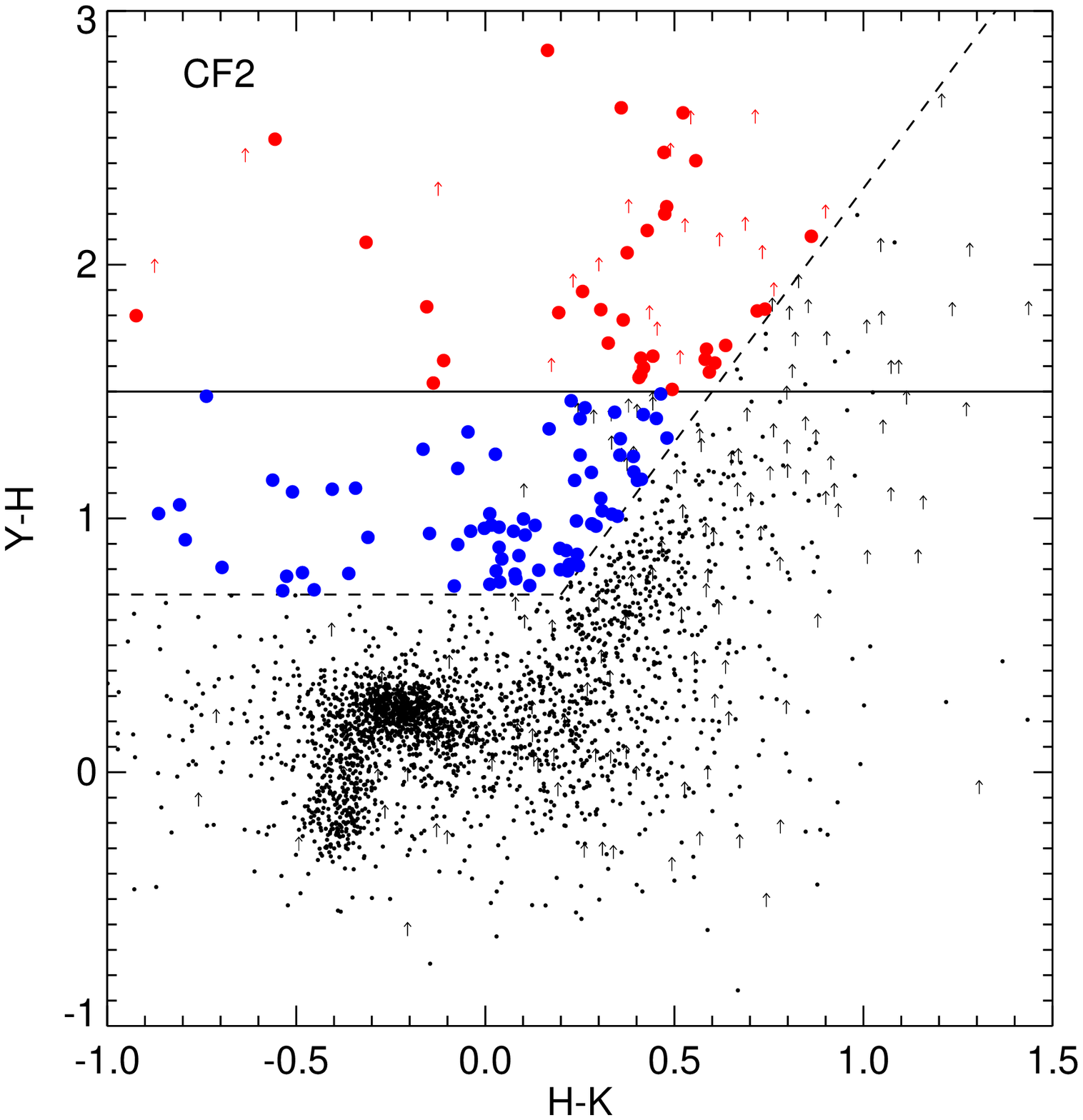}
\includegraphics[width=5.7cm,angle=0,bb=0 10 500 600]{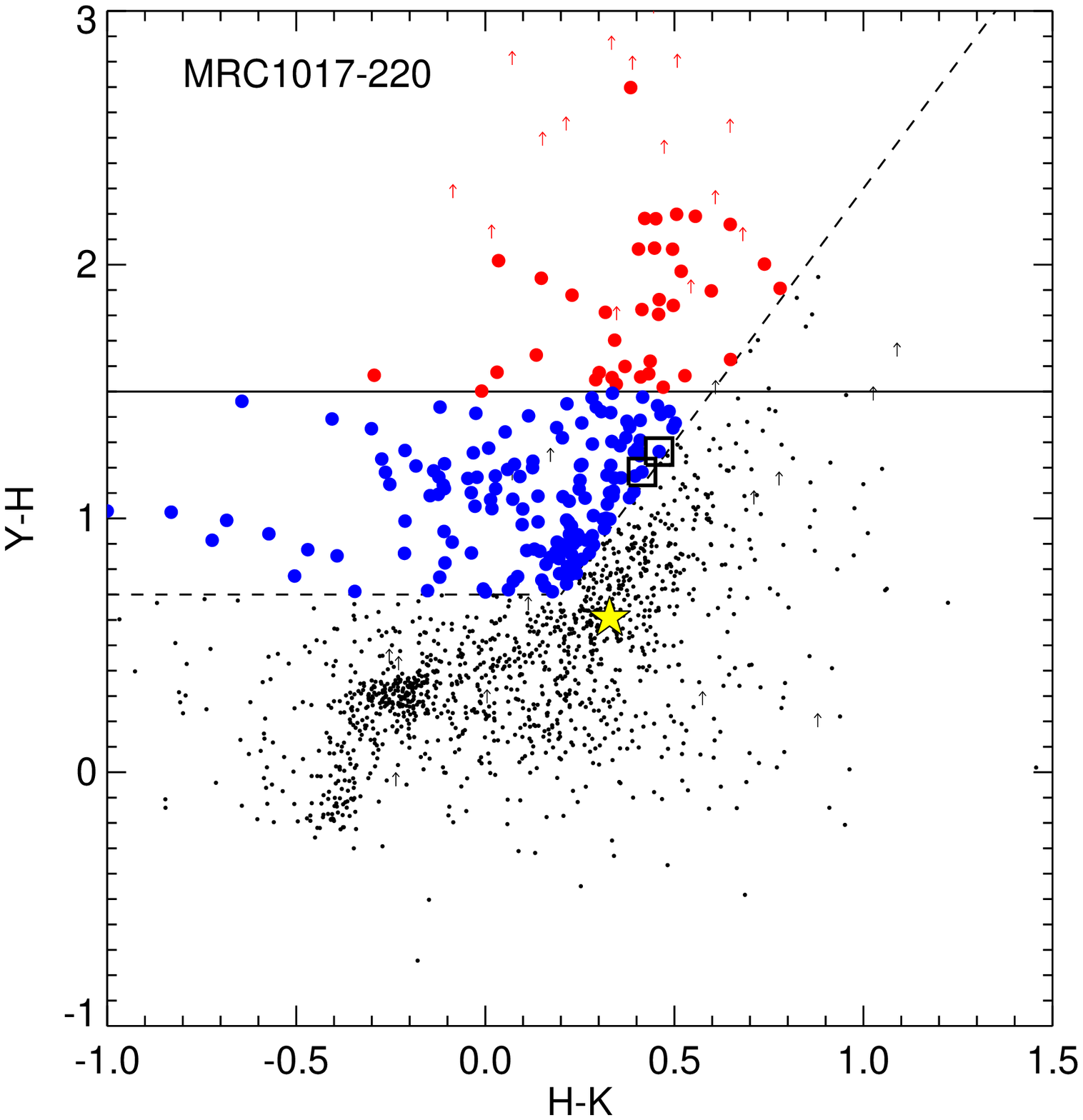}
\includegraphics[width=5.7cm,angle=0,bb=0 10 500 600]{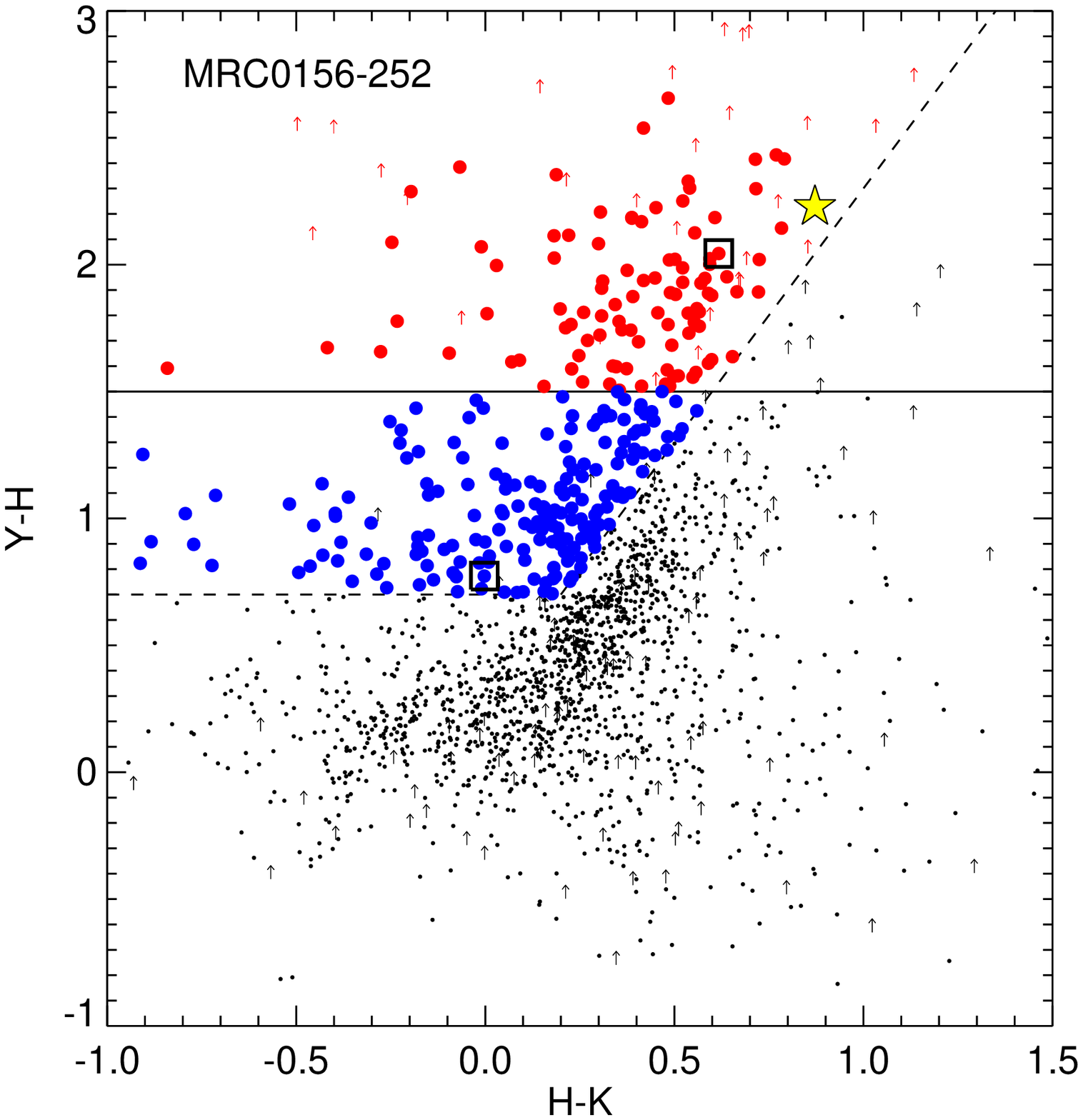}
\end{center}
\caption{Colour-colour diagrams $Y-H$ vs $H-K$ for all the fields : {\it Top panels:} 
Control fields: GOODS1 (left), GOODS2 (middle) and CF2 (right). {\it Bottom panels:} 
HzRG fields: MRC~1017-220 (left) and MRC~0156-252 (right). We plot galaxies detected 
down to the $2\sigma$ detection limits in $Y$, $H$ and $Ks$ (black dots). r-$YHK$ 
and b-$YHK$ galaxies are designed by red and blue circles. Arrows indicate lower limits for
sources detected in $H$ and $Ks$ but not in $Y$.
The radio galaxies are marked by the yellow stars in the bottom panels. 
We also indicate with black open squares the two EROs found in the field of MRC~1017-220 
(bottom left panel, Cimatti et al.~1999) and the two sources found within $5\arcsec$ of MRC~0156-252 
(bottom right panel) corresponding to objects B (the b-$YHK$ galaxy) and C (the r-$YHK$ galaxy) 
in Fig.~\ref{zoom}.}
\label{CCD}
\end{figure*}

We further test the $YHK$ criterion using the photometric redshifts ($z_{phot}$; hereafter) 
available in the GOODS-MUSIC catalogue. Using the multiwavelength photometry of the 
GOODS-S field, \citet{Grazian2006B} applied a photometric redshift code to their catalogue. 
They tested their code with the available spectroscopic redshifts and found an accuracy of 
$\sigma_{z} = 0.03\times(1+z)$ \citep{Grazian2006A}. 

We examine the $z_{phot}$ of the $YHK$-selected galaxies within 
the $2\sigma$ detection limits as well as within the 
$90$\% completeness limits of the HAWK-I data i.e., the 
completeness of passive red ellipticals for the r-$YHK$ galaxies and the completeness of spiral 
galaxies for the b-$YHK$ galaxies. Fig.~\ref{zsel} shows the photometric redshift distribution 
of the r-$YHK$ galaxies (top panel) and the b-$YHK$ galaxies (bottom panel). We find 
that $77$\% ($2\sigma$; $71$\% in the completeness limits) of the r-$YHK$ galaxies and 
$68$\% ($47$\%) of the b-$YHK$ have $z_{phot}>1.6$ which confirms the efficiency of the 
selection criteria. 

The r-$YHK$ criterion is efficient at selecting galaxies at the targeted redshifts 
with $42$\% ($50$\%) of the sources with $1.6<z_{phot}<2.1$. Within the limits of completeness, 
the b-$YHK$ criterion is also efficient with $38$\% of galaxies having $1.6<z_{phot}<2.1$. The 
photometric redshift distribution becomes much broader when considering fainter sources and 
extends towards higher redshift sources. At the $2\sigma$ limits, $27$\% of b-$YHK$ sources 
have $1.6<z<2.1$. We note that if indeed an overdensity of red or blue galaxies is present in the 
surroundings of the HzRGs, the sample of $YHK$-selected galaxies in these fields would 
contain a higher fraction of sources at the targeted redshifts. The percentages given earlier are 
therefore expected to be lower limits.

\subsection{The $YHK$-selected galaxies}

We apply the $YHK$ criteria to the five fields with $YHK$ coverage i.e., MRC~1017-220, 
MRC~0156-252, CF2, GOODS1 and GOODS2. We consider sources with a $2\sigma$ detection
in all three bands. The $Y-H$ vs $H-K$ colour-colour diagrams for the five fields are shown
in Fig.~\ref{CCD}. Red and blue circles indicate r-$YHK$ and b-$YHK$ galaxies respectively.

We overplot sources detected (within $2\sigma$ limits) in $H$ and $Ks$ but not in our 
$Y$-band catalogue. In order to place these sources in the colour-colour diagram, we assume
a lower limit on the $Y$-band magnitude (see arrows, Fig.~\ref{CCD}). We use SExtractor in dual 
mode using the source positions in the $Ks$-band and deriving aperture photometry for these 
sources on the $Y$-band. For sources brighter that our $2\sigma$ detection limits but that were 
not part of our $Y$-band catalogue (i.e.~beyond our completeness limits), we assign the $Y$-band 
photometry derived from the aperture placed at  the $Ks$-band source position. For fainter sources, 
we assign them the $2\sigma$ limits of our $Y$-band images. Sources with lower limits falling in the 
r-$YHK$ selection area of the colour-colour diagram are overplotted in red. These objects are 
particularly interesting since they have very red $Y-H$ colours which suggests they have 
strong $4000$\AA~breaks.

Within our $2\sigma$ limits in $Y$, $H$ and $Ks$, we find $38$ r-$YHK$ galaxies ($151$ 
b-$YHK$ galaxies) in MRC~1017-220, $105$ ($191$) in MRC~0156-252, $47$ ($196$) in 
GOODS1, $48$ ($176$) in GOODS2 and $38$ ($81$) in CF2. We note that since the $2\sigma$ 
magnitude limits and area slightly vary from field to field, these numbers are not directly 
comparable to one another. 

\citet{Cimatti2000} looked at the populations of EROs ($(R-K)_{Vega}>6$) in $14$ fields around 
radio-loud AGN at $z>1.5$, including MRC~1017-220. An excess of EROs was found in the 
field of MRC~1017-220, with three EROs located within $2.5\arcmin$ of the HzRG.
Two of the three were observed in spectroscopy in the $H$-band with VLT/ISAAC  
\citep{Cimatti1999} and a `spectro-photometric' redshift  ($z_{sphot}$) was derived for 
both sources combining the ISAAC spectrum continuum and broad-band photometry. 
The two sources, J101948-2219.8 
(R.A.: 10:19:47.79, Dec.: -22:19:46.6, $K_{Vega}=18.7$, $z_{sphot}=1.52 \pm 0.12$) 
and J101950-2220.9
(R.A.: 10:19:49.76, Dec.: -22:20:53.9, $K_{Vega}=18.6$, $z_{sphot}=1.50 \pm 0.25$), 
were both classified as early-type galaxies due to their SEDs consistent with no dust extinction. We 
look at the colours of these two objects 
(see black squares, Fig.~\ref{CCD}, bottom left panel) and both are selected by the b-$YHK$ criterion.
However, only deeper spectroscopy over a wider wavelength range will confirm whether those 
two targets have redshifts consistent with MRC~1017-220.

\begin{figure} 
\begin{center} 
\includegraphics[width=7.7cm,angle=0]{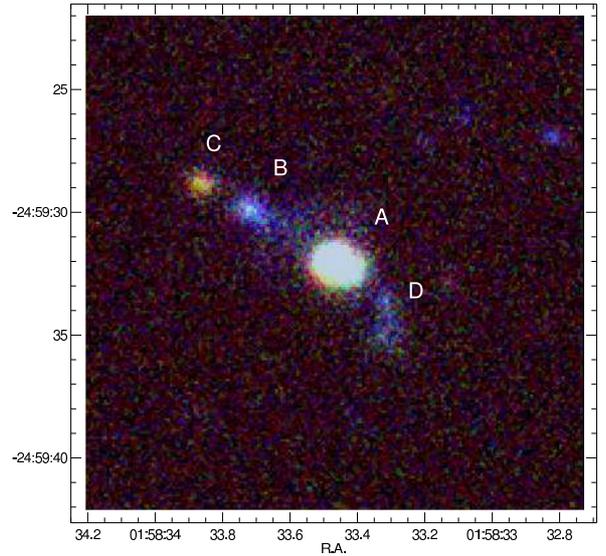}
\end{center}
\caption{$3$-colour image (R, G, B for $Ks$, $H$, $Y$) of the $20\arcsec \times 20\arcsec$
field of view around MRC~0156-252 (North is Up, East Left). A is the radio galaxy. 
B and C, previously reported in Pentericci et al.~2001, are respectively selected by the r-$YHK$ 
and b-$YHK$ criterion. An additional component, D, is also found aligned with MRC~1056-252, B and
C but too faint in $Ks$ to be classified by our near-infrared criterion.}
\label{zoom}
\end{figure}

\citet{Pentericci2001} presented near-infrared NICMOS imaging of MRC~0156-252. They found 
two objects within $5\arcsec$ of the radio galaxy, the three sources being aligned in the direction 
of the radio axis (i.e.~NE/SW). Fig.~\ref{zoom} is a three-colour image of the close vicinity of MRC~0156-252. 
Three objects are observed near MRC~0156-252. We label the components using the same 
notation as \citet{Pentericci2001} (A for MRC~0156-252, B and C for the two eastern 
components), adding `D' for the third faint source to the west of the HzRG. As previously noticed 
in \citet{Pentericci2001}, component C is much redder than component B. C is a r-$YHK$ 
galaxy and B is a b-$YHK$ galaxy. B and C are indicated by black 
squares in Fig.~\ref{CCD}, bottom right panel. D is detected but is very faint in $Ks$ ($<2\sigma$ level) 
and was therefore not considered in the candidate selection process. Spectroscopy is required to 
prove the association of these objects with the radio galaxy but their physical closeness and 
colours strongly suggest that they are associated with MRC~0156-252.
The scale of this system is similar to the structure associated with PKS~1138-262 i.e.,~about 
$15\arcsec$ \citep[see][Fig.~2]{Miley2006}. 

\section{Candidates properties}

\subsection{Surface densities of $YHK$-selected galaxies}

We compare the surface densities of the $YHK$-selected galaxies in the five fields to 
determine whether there is an overdensity of sources around the targeted HzRGs. 
For a direct field to field comparison, we cut the catalogues at the completeness 
limits of the shallowest HAWK-I images (see \S3; $Y<24$, $H<22.9$
and $Ks<22.4$). Table~\ref{densities} summarizes
the densities for both the r-$YHK$ and b-$YHK$ galaxies in the five fields assuming 
Poisson errors for the source densities. 

As far as the r-$YHK$ galaxies are concerned, the fields around MRC~1017-220, 
CF2 and GOODS1 have similar densities. As expected, GOODS2 which contains 
the overdensity at $z=1.6$ described in \citet{Kurk2009} is slightly denser than average, 
e.g.~by a factor of $1.7 \pm 0.7$ compared to GOODS1. The field around MRC~0156-252 
is significantly overdense compared to all the other fields: by a factor of $3.1 \pm 1.1$ 
compared to MRC~1017-220, $3.5 \pm 1.3$ compared to GOODS1 and $2.1 \pm 0.6$
compared to GOODS2\footnote[6]{The $YHK$ colour selection is very
sensitive to the photometric calibration of the images and the $Y$-band zeropoint has relatively 
large error bars ($0.07$). We note however that even if the zeropoint was offset by 
-0.07, the field would still be overdense in red galaxies by a factor of $2.4$ compared to MRC~1017-220 
and $1.6$ compared to GOODS2.}. 

We also derive densities within $1$~Mpc ($\sim 2\arcmin$, angular separation) for
both HzRGs corresponding to the classical estimates
of virial radius in the highest redshift clusters known to date \citep[e.g. ][]{Hilton2007}.
In the close vicinity of MRC~0156-252, the r-$YHK$ density is even higher: $3.9 \pm 1.9$ times 
denser than GOODS1 and $2.4 \pm 1.1$ times denser than the (overdense) GOODS2 field.

As far as the b-$YHK$ galaxies are concerned, densities are more similar from field
to field. However, we find that GOODS2 and the HzRGs fields are slightly denser 
than average. MRC~1017-220 is the densest field. The region within $1$~Mpc 
of the radio galaxy is $2.2 \pm 0.8$ times denser than CF2. The field of MRC~0156-252 
is also denser, by a factor of $1.4 \pm 0.4$ compared to CF2.

\begin{table}
\caption{Number densities of $YHK$-selected galaxies}
\centering
\begin{tabular}{l c c c}
Field & Area & r-$YHK$$^{\mathrm{a}}$ & b-$YHK$$^{\mathrm{a}}$ \\ 
 & (arcmin$^2$) & (deg$^2$) & (deg$^2$) \\ 
 \hline
MRC1017-220				&	$54.9$	&	$720 \pm220$		&	$2530 \pm 410$	\\
MRC1017-220 ($<1$Mpc) 	&	$11.0$	&	$1800 \pm 770$	&	$3240 \pm 1030$	\\
MRC0156-252				&	$59.0$	&	$2220 \pm 370$	&	$2010 \pm 350$	\\
MRC0156-252 ($<1$Mpc)	&	$11.0$	&	$2520 \pm 910$	&	$1800 \pm 770$	\\
CF2						&	$54.3$	&	$930 \pm 250$		&	$1460 \pm 310$	\\
GOODS1					&	$50.3$	&	$640 \pm 210$		&	$1570 \pm 340$	\\
GOODS2					&	$50.3$	&	$1070 \pm 280$	&	$1790 \pm 360$	\\
\hline
\end{tabular}
\begin{list}{}{}
\item[$^{\mathrm{a}}$] Within $90$\% completeness limits; densities were rounded for clarity. 
\end{list}
\label{densities}
\end{table}

\begin{figure}
\begin{center}
\includegraphics[width=7cm,angle=0]{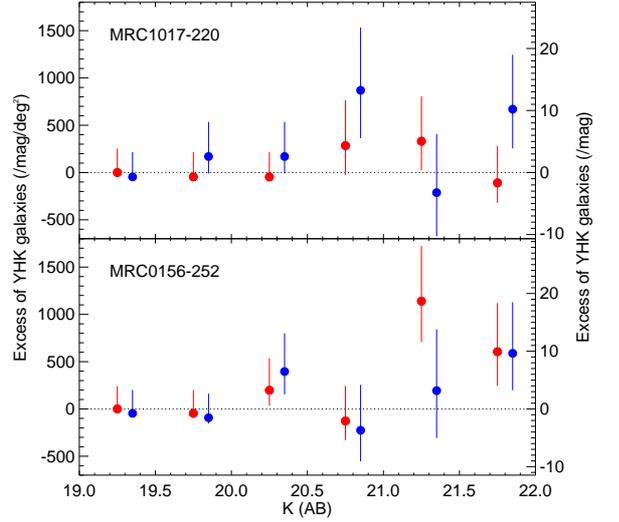}
\end{center}
\caption{Number counts of the excess of $YHK$-selected galaxies (per $0.5$~$Ks$~mag bin) 
found in the surroundings of MRC~1017-220 (top panel) and MRC~0156-252 (bottom panel). 
We first derive reference number counts for 
the r-$YHK$ (red) and b-$YHK$ (blue) galaxies from the control fields: CF2+GOODS1+GOODS2 
and then subtract them from the number counts of the $YHK$ galaxies found in the field of 
the HzRGs. We assume poissonian errors. Densities are given in sq. deg. (left axis). Excess number
of galaxies for each HzRG field ($\sim 60$ sq. arcmin) is shown on the right axis.}
\label{nc}
\end{figure}

\begin{figure*}[!ht] 
\begin{center} 
\includegraphics[width=9.1cm,angle=0,bb=0 30 500 580]{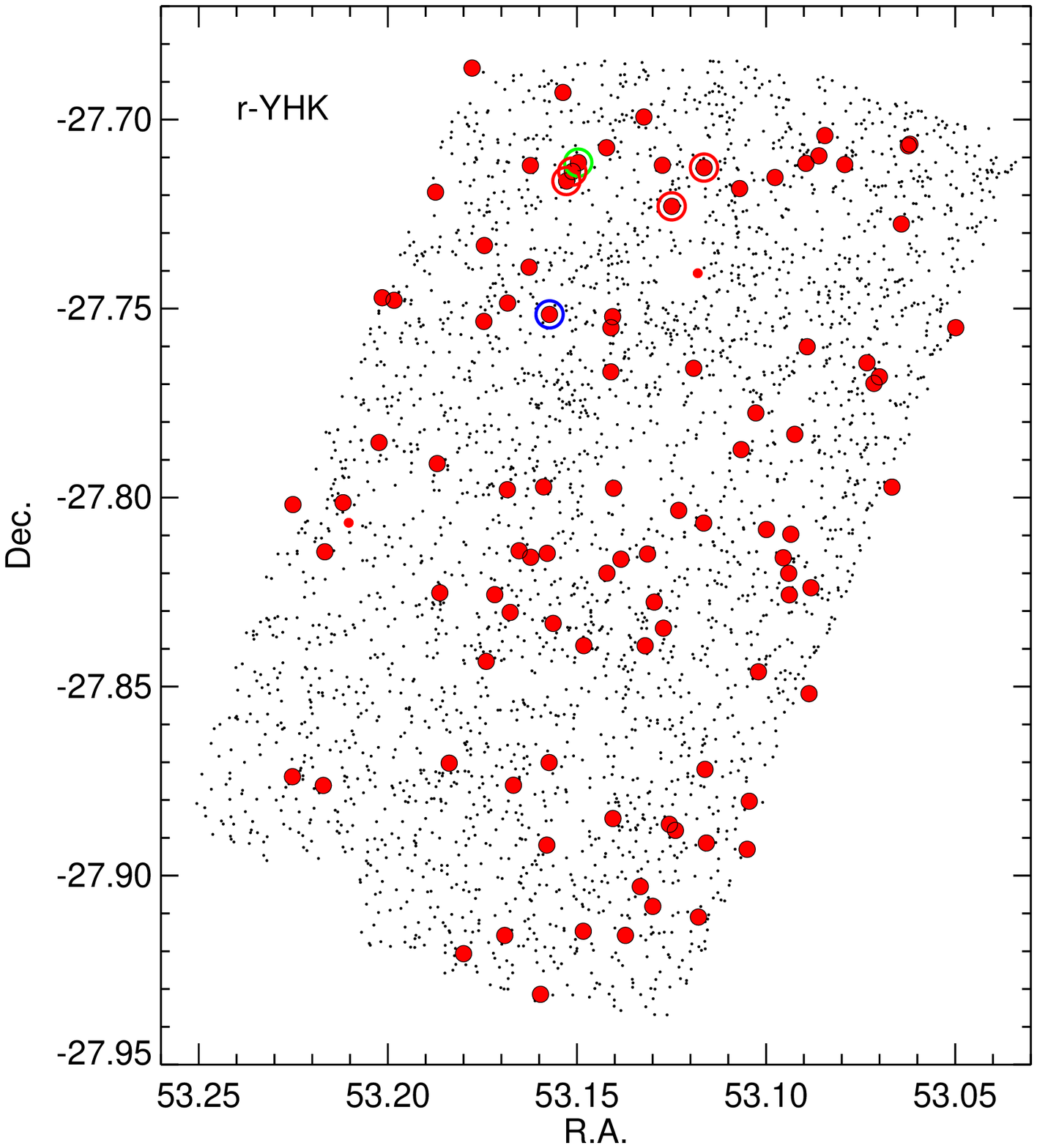}
\includegraphics[width=9.1cm,angle=0,bb=0 30 500 580]{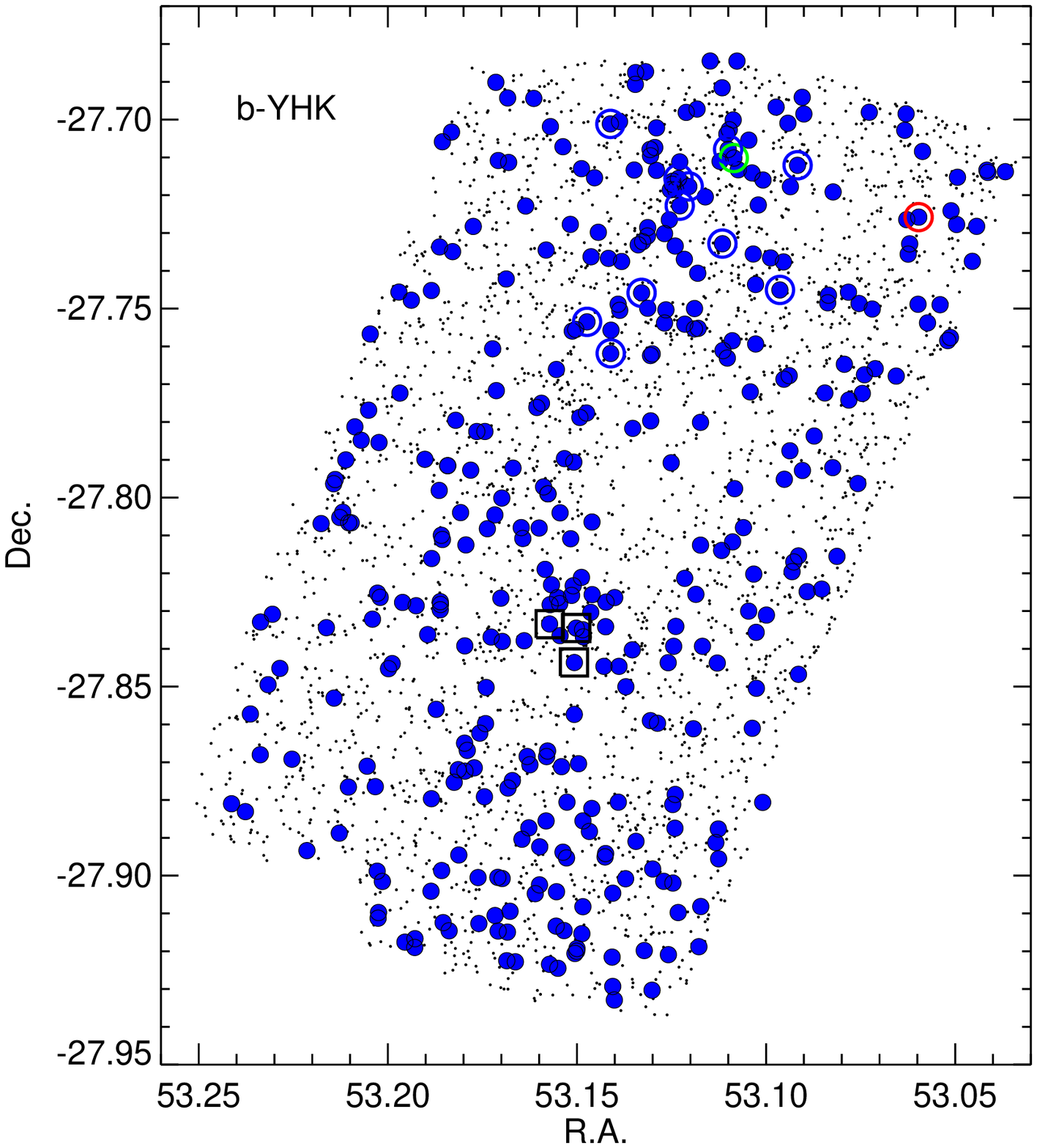}
\end{center}
\caption{Spatial distribution of the $YHK$-selected galaxies in the GOODS fields 
(GOODS1+GOODS2). We also indicate with open circles 
the $6$ r-$YHK$ and $12$ b-$YHK$ galaxies, spectroscopically confirmed members of the overdensity at 
$z\sim1.6$ reported by \citet{Kurk2009}, symbols colours accounting for the object 
spectral class (see Fig.~\ref{GOODSzspec}, lower panel). Three b-$YHK$ galaxies 
confirmed at $z\sim1.6$ from the ESO GOODS spectroscopy are also indicated by black squares.}
\label{radec_goods}
\end{figure*}

We also derive the number counts of the $YHK$-selected galaxies in all the targeted fields. Reference number
counts are derived for both r-$YHK$ and b-$YHK$ selected galaxies combining CF2, GOODS1, GOODS2 and 
subtracted from the number counts of the targeted HzRGs fields.
Fig.~\ref{nc} illustrates the number counts of this `excess' of $YHK$ galaxies per $0.5$~$Ks$~mag bin suspected to
be associated with the HzRGs. Errors on both the radio galaxies fields and reference fields number counts
are added in quadrature. We assume the \citet{Gehrels1986} small numbers approximation for Poisson distributions.

As for Table.\ref{densities}, this analysis is made in the limits of completeness
of the (shallowest) data for a direct field to field comparison. However, due to the selection technique combined with
our completeness limits, we do not select all $YHK$ galaxies in the fields with the selection getting rapidly incomplete
at fainter magnitudes. To illustrate this, we look at the colours of r-$YHK$ galaxies in the field of MRC~0156-252. On 
average, they have $\langle H-Ks \rangle \sim 0.4$ and $\langle Y-H \rangle \sim 1.7$. The completeness limit in $Y$
(our limiting band) is $Y=24$ corresponding to $H=22.3$ and therefore $Ks=21.9$, despite the data being complete
up to half a magnitude deeper in $Ks$. We therefore limit the analysis to $Ks<22$. 

No significant excess is
seen in the r-$YHK$ number counts of the MRC~1017-220 field compared to the control fields. 
The excess of b-$YHK$, already mentioned earlier in this section is also observed in Fig.~\ref{nc} 
with an excess of $Ks>20.5$ sources. The field of MRC~0156-252 also shows an excess of blue 
$Ks>20$ sources. We do not observe an excess of bright red sources ($Ks<20$) in the field of 
MRC~0156-252 as compared to the control fields. The overdensity of red sources in the field of 
MRC~0156-252 becomes prominent at $Ks>21$ with an excess of $10$ to $20$ galaxies in the 
field (by bin of $0.5$mag) compared to average. At $z\sim2$, such magnitudes in $Ks$ for elliptical 
galaxies correspond to masses of several $10^{11}$M$_{\odot}$ \citep{Kodama2007, Rocca2004} 
suggesting that if the overdensity detected is indeed associated with MRC~0156-252, the HzRG 
would lie in a structure that already contains very massive, passively evolving galaxies.

\begin{figure*}[!ht] 
\begin{center} 
\includegraphics[width=7.5cm,angle=0]{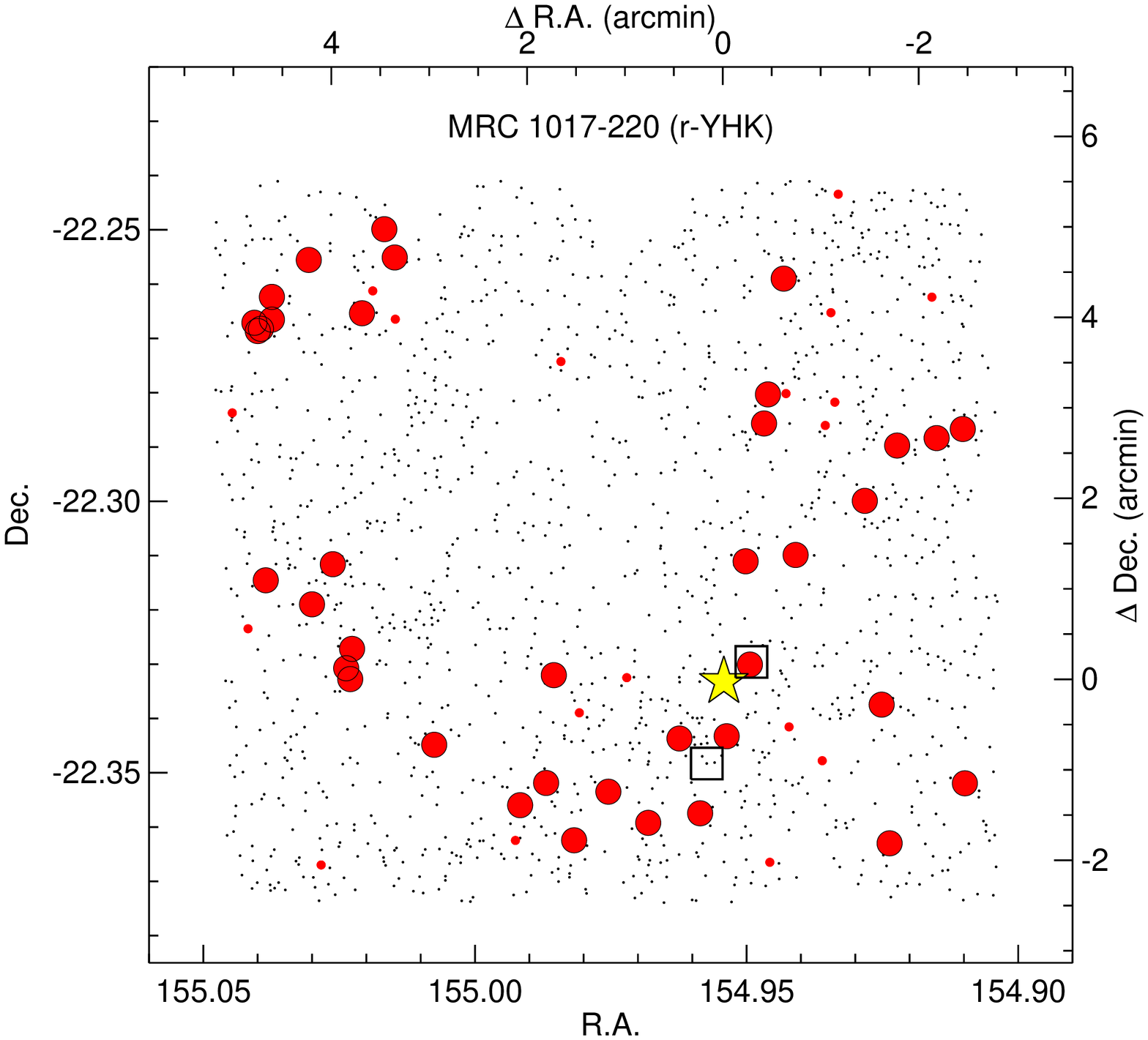}
\includegraphics[width=7.5cm,angle=0]{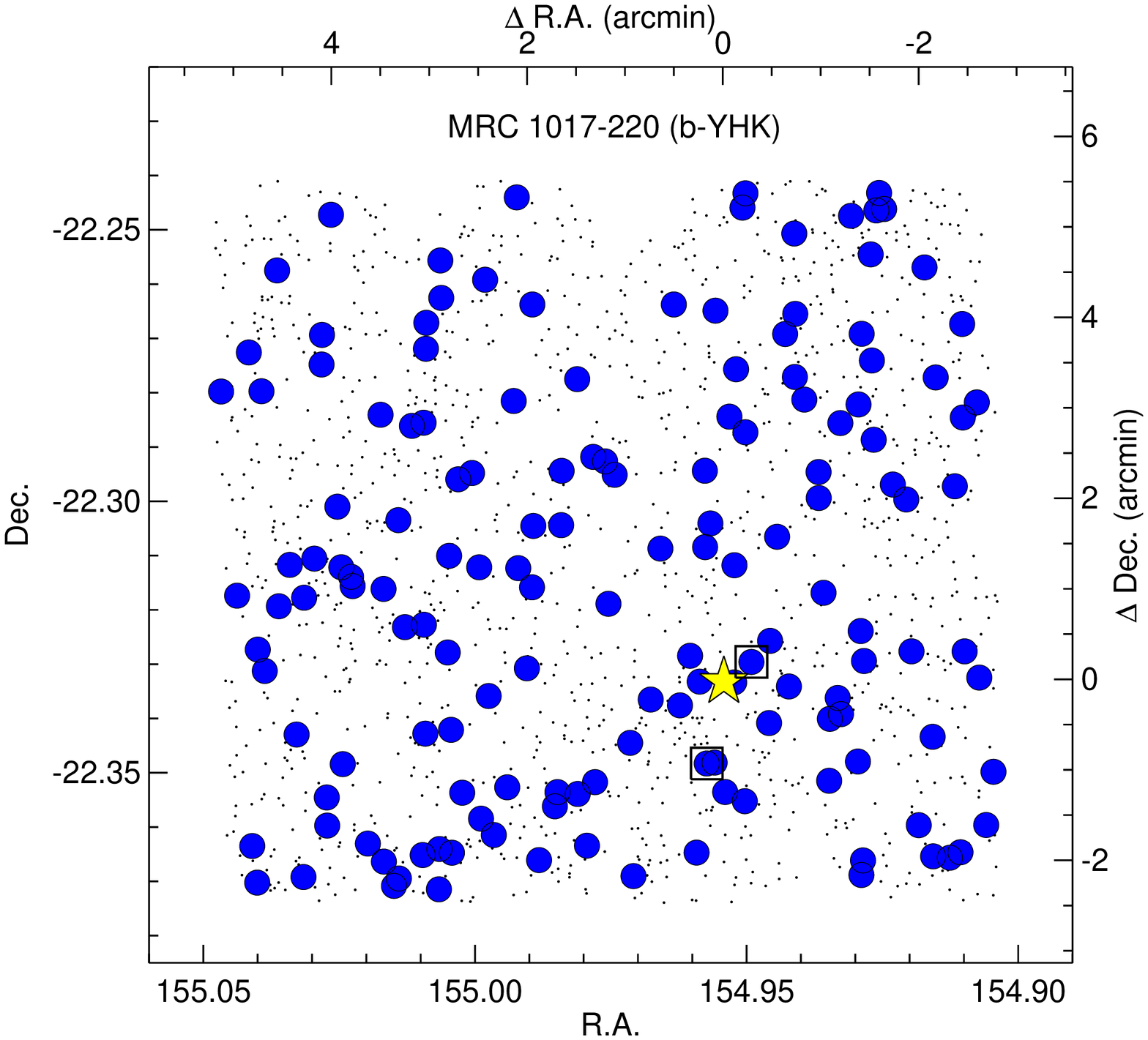}
\includegraphics[width=7.5cm,angle=0]{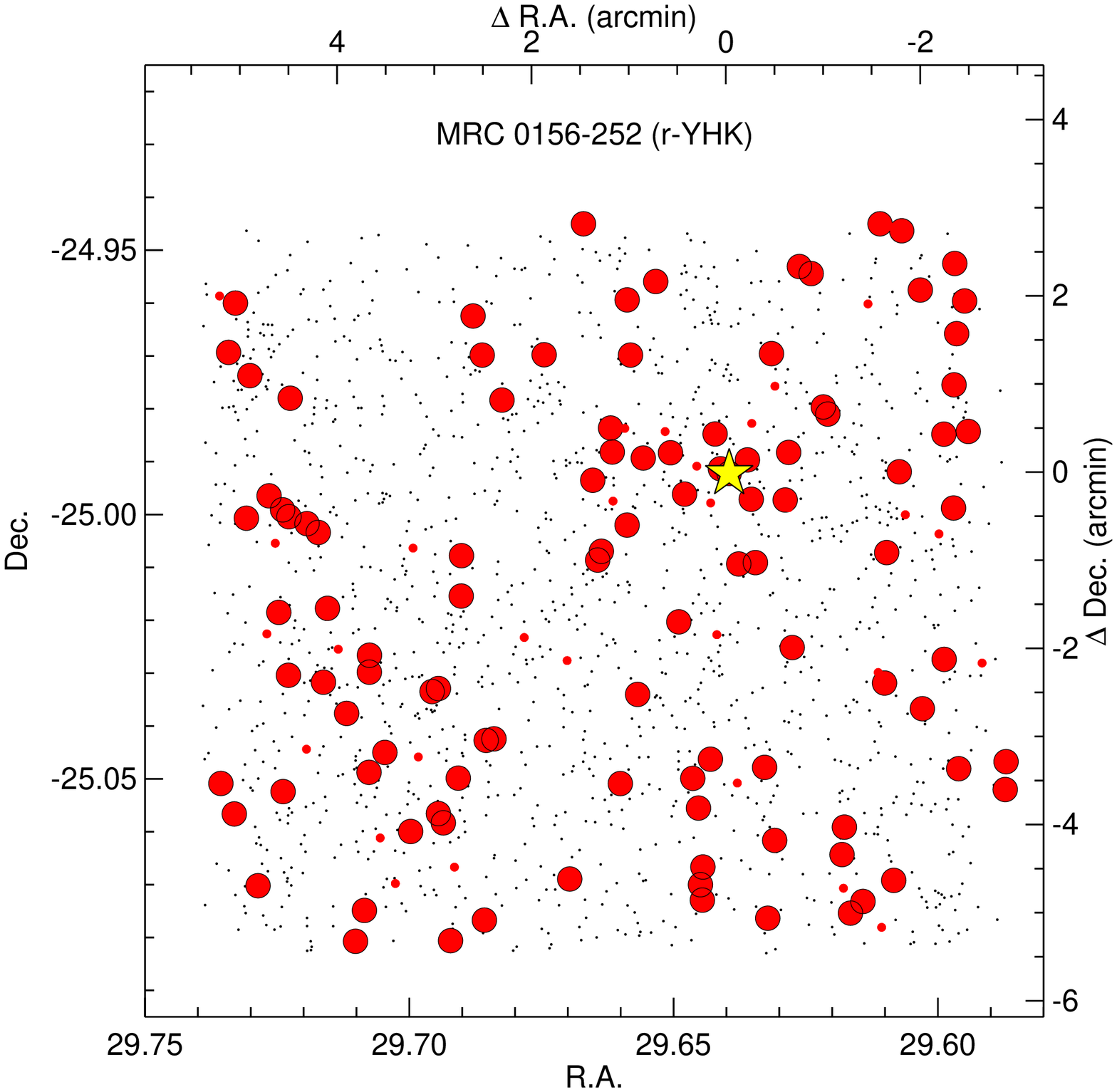}
\includegraphics[width=7.5cm,angle=0]{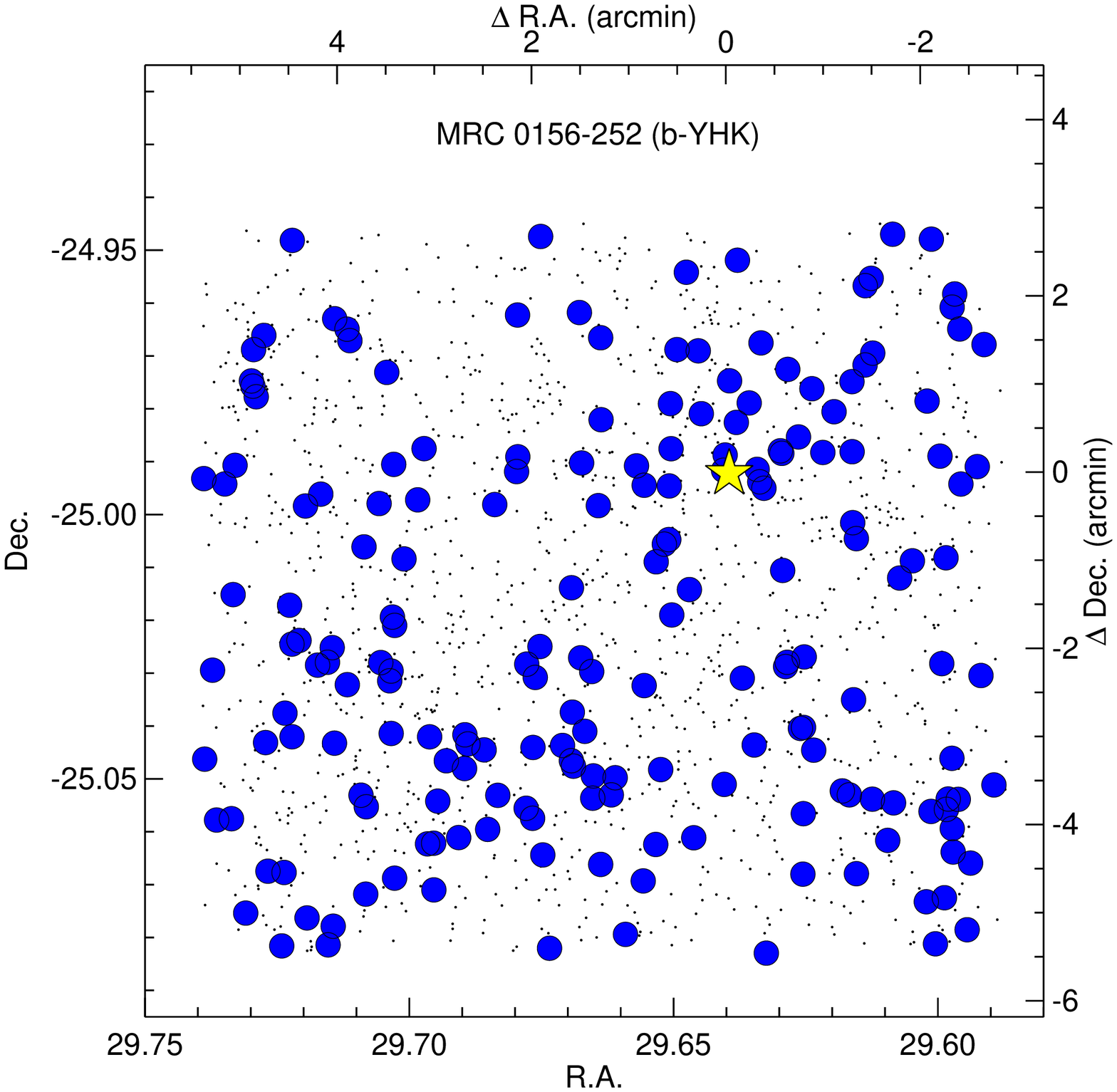}
\includegraphics[width=7.5cm,angle=0]{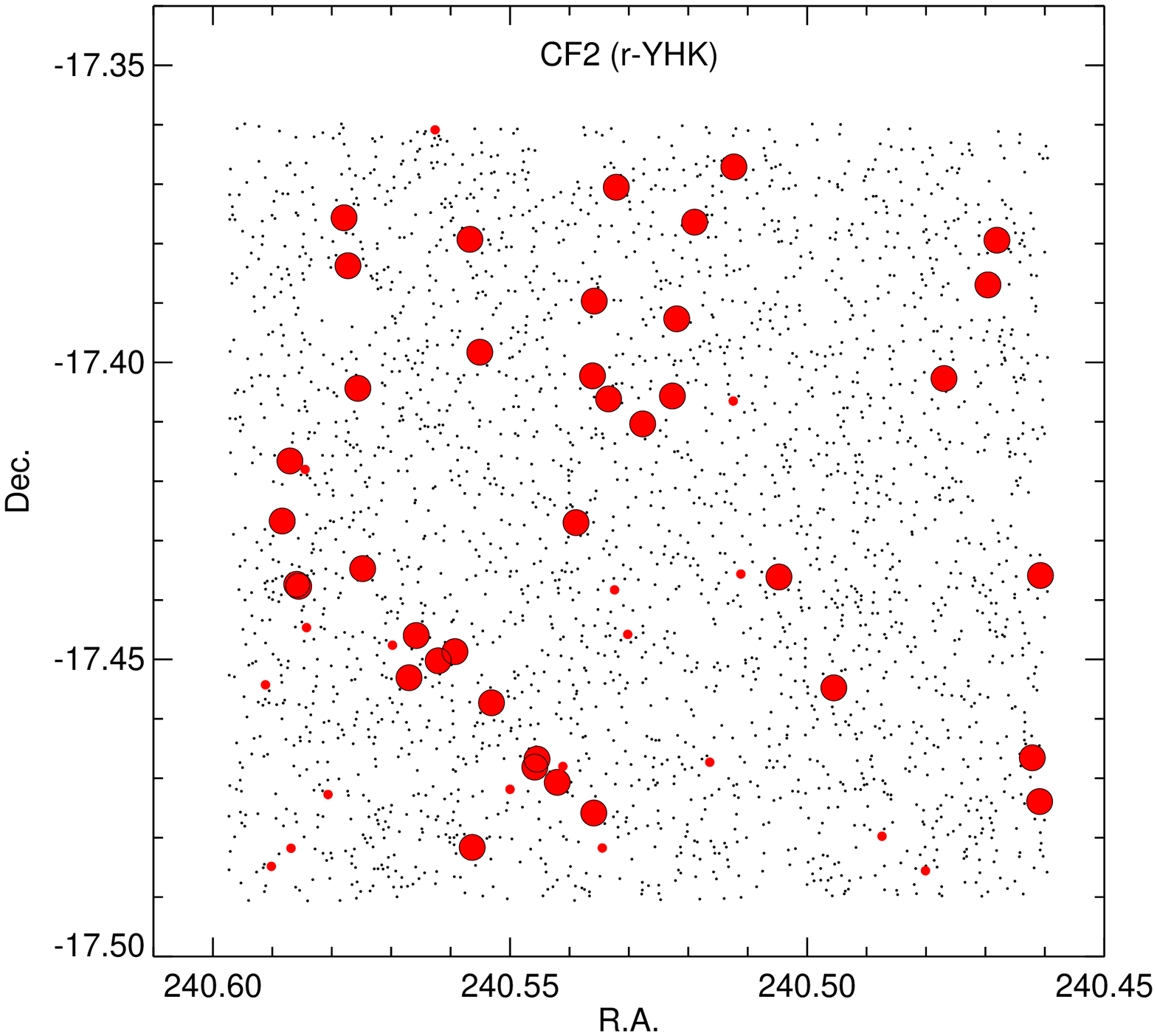}
\includegraphics[width=7.5cm,angle=0]{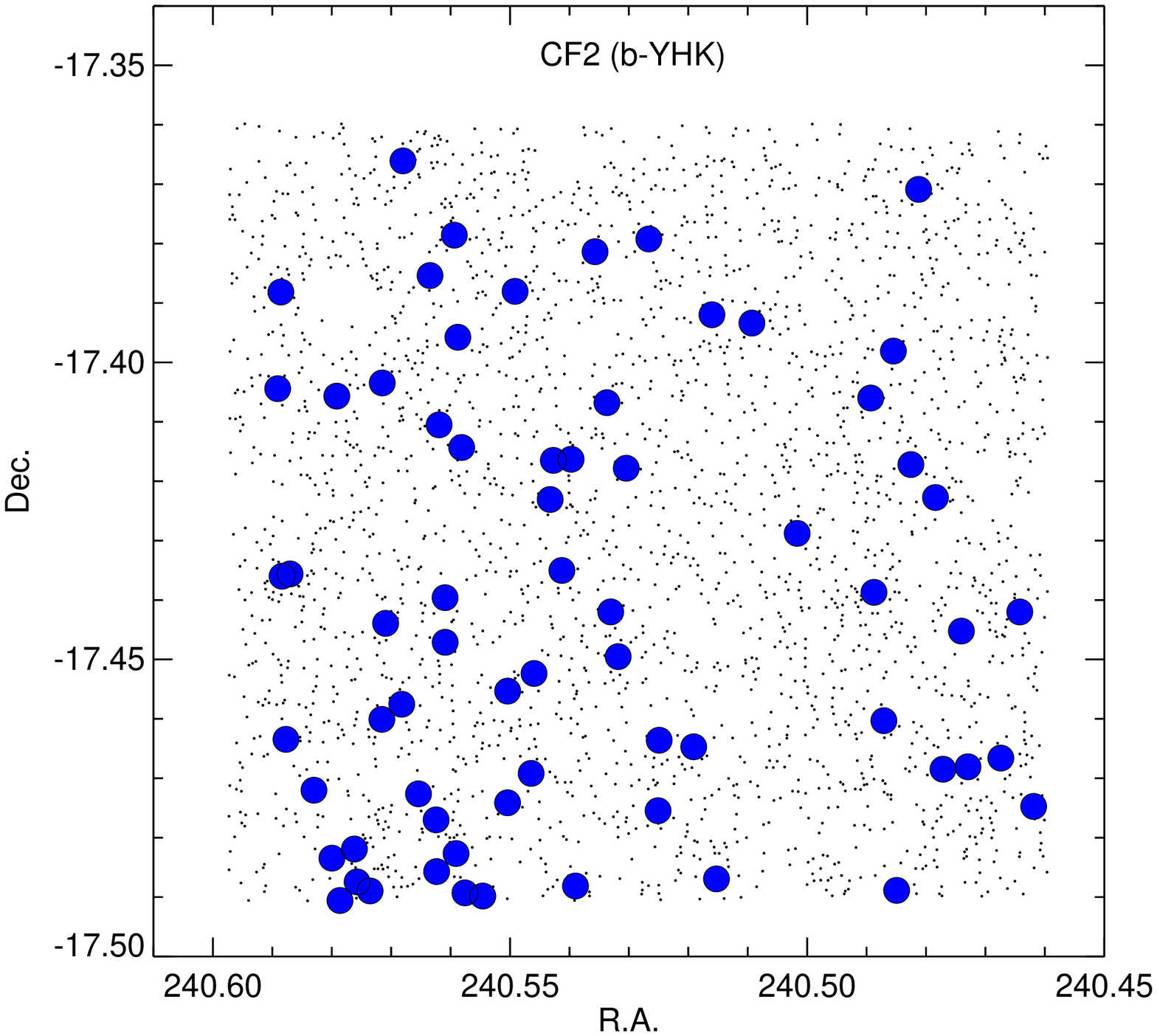}
\end{center}
\caption{Spatial distribution of the $YHK$-selected galaxies in the fields around MRC~1017-220 and 
MRC~0156-252 and the control field CF2: b-$YHK$ galaxies in blue and r-$YHK$ galaxies in red (arrows
of Fig.~\ref{CCD} in small red dots).} 
\label{radec}
\end{figure*}

\subsection{Spatial distribution of the candidates}

The spatial distribution of the $YHK$-selected galaxies (detected within our $2\sigma$ 
magnitude limits) in the two GOODS fields is 
shown in Fig.~\ref{radec_goods} for r-$YHK$ and b-$YHK$ galaxies in 
red and blue large circles. Sources detected in $Y$, $H$ and $Ks$ ($2\sigma$) are 
marked with black dots. Sources detected in $H$ and $Ks$ but not in $Y$, though
falling in the r-$YHK$ selection area (i.e.~red arrows in Fig.~\ref{CCD}), are overplotted in small red dots.
Due to the deep GOODS $Y$-band, there are only two such sources.
We observe a clear inhomogeneity in the distributions of r-$YHK$ and b-$YHK$ 
galaxies in  the northern part of the field (GOODS2). A concentration of both red and blue 
galaxies is found at the location of the $z=1.6$ galaxy overdensity presented in 
\citet{Kurk2009}, confirming again the efficiency of the $YHK$ criterion at detecting
$z\geq1.6$ galaxy structures.

We note that the northern part of GOODS1 (the southern part of the GOODS field) 
is more populated with r-$YHK$ galaxies than the rest of GOODS1. The spatial distribution of b-$YHK$ galaxies 
is more homogenous, but we also observe a clear concentration of b-$YHK$ galaxies at the same 
position as the excess of r-$YHK$ galaxies (i.e.~at R.A.$\sim53.15$ and Dec.$\sim-27.83$). 
We therefore confirm the results of \citet{Kurk2009} and \citet{Castellano2007} that the overdensity
spreads southward.
Three of the b-$YHK$ galaxies have a spectroscopically 
confirmed redshift of $\sim1.6$ from the ESO GOODS spectroscopy (reported in the 
GOODS-MUSIC catalogue; see the three black squares in Fig.~\ref{radec_goods}, right panel).

The spatial distribution of the $YHK$ galaxies around MRC~1017-220,
MRC~0156-252 and in CF2 is shown in Fig.~\ref{radec}. r-$YHK$ 
and b-$YHK$ galaxies are shown as red and blue large circles (first and second column) 
respectively.
Sources detected in $H$ and $Ks$ with lower limits in $Y$ and selected by our
r-$YHK$ criteria are shown by small red dots.
Distances from both HzRGs (yellow star) are indicated on the top and right axis 
of the four first panels. We do not see any specific distribution of $YHK$-selected 
galaxies in CF2 (Fig.~\ref{radec}, bottom row). 

For MRC~1017-220, we indicate the two EROs with 
spectro-photometric redshift from \citet{Cimatti1999} by black squares. 
The r-$YHK$-selected galaxies have a non uniform spatial distribution over the field,
with a hint of a filamentary distribution in the NW-SE direction in which lies the radio galaxy.
No clear spatial inhomogeneity is observed for the b-$YHK$ galaxies except a slight excess
in the southern part of the field.
We note that this result is in agreement with previous studies which show that red galaxies
are more strongly clustered than blue ones \citep{Daddi2000, Brown2003, Kong2006}.

\begin{figure*}
\begin{center}
\includegraphics[width=8cm,angle=0]{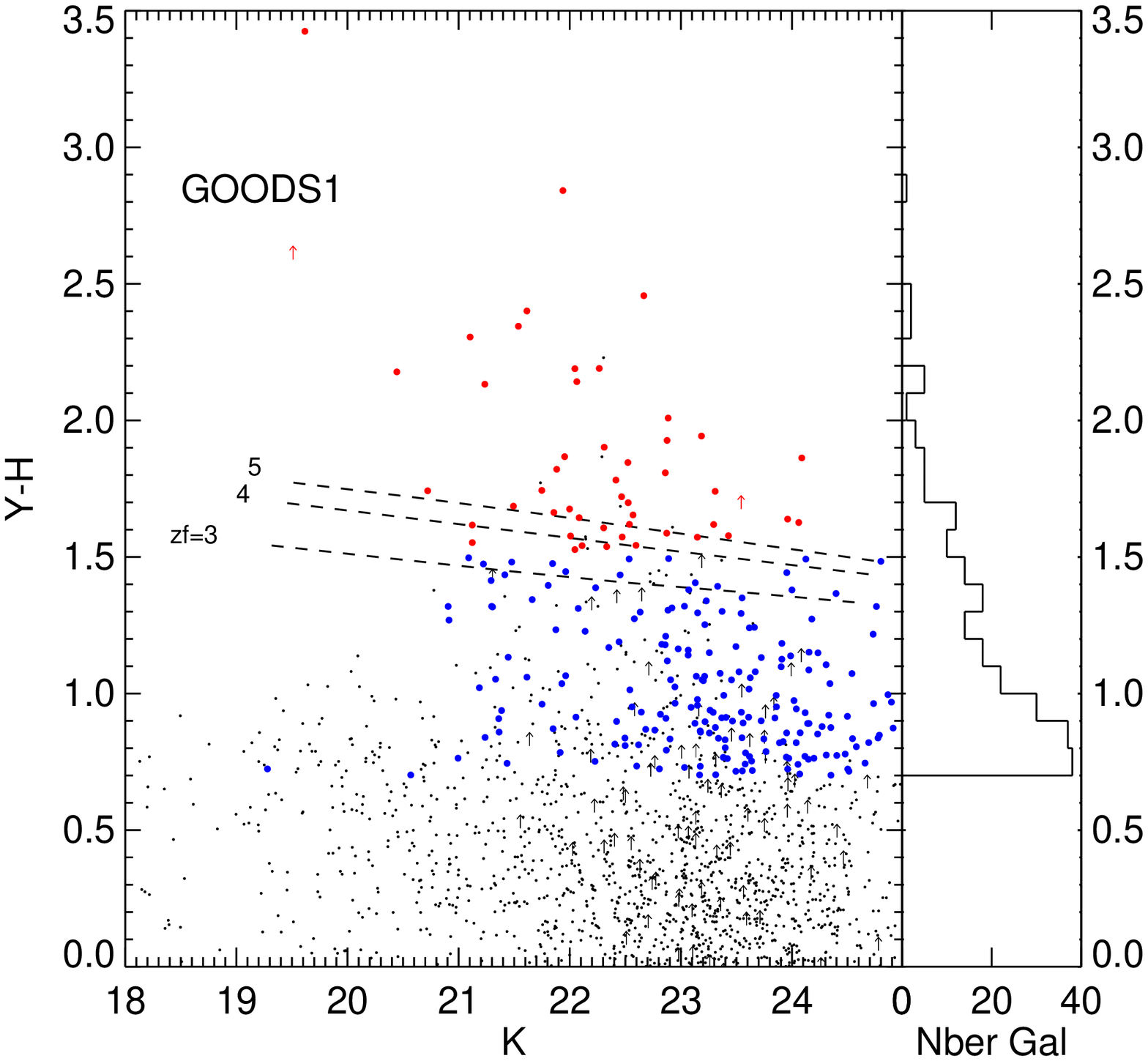}
\includegraphics[width=8cm,angle=0]{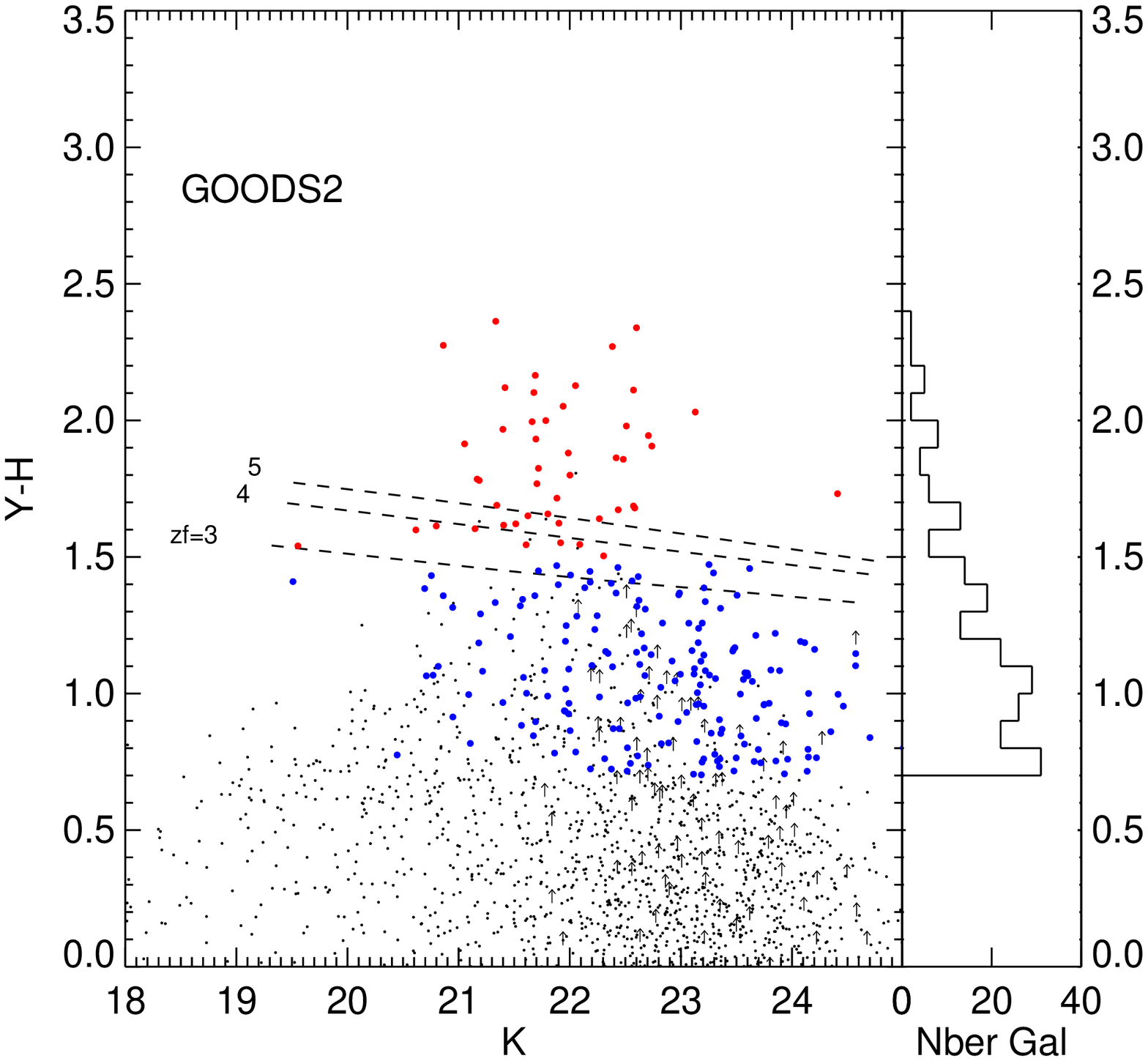}\\
\includegraphics[width=8cm,angle=0]{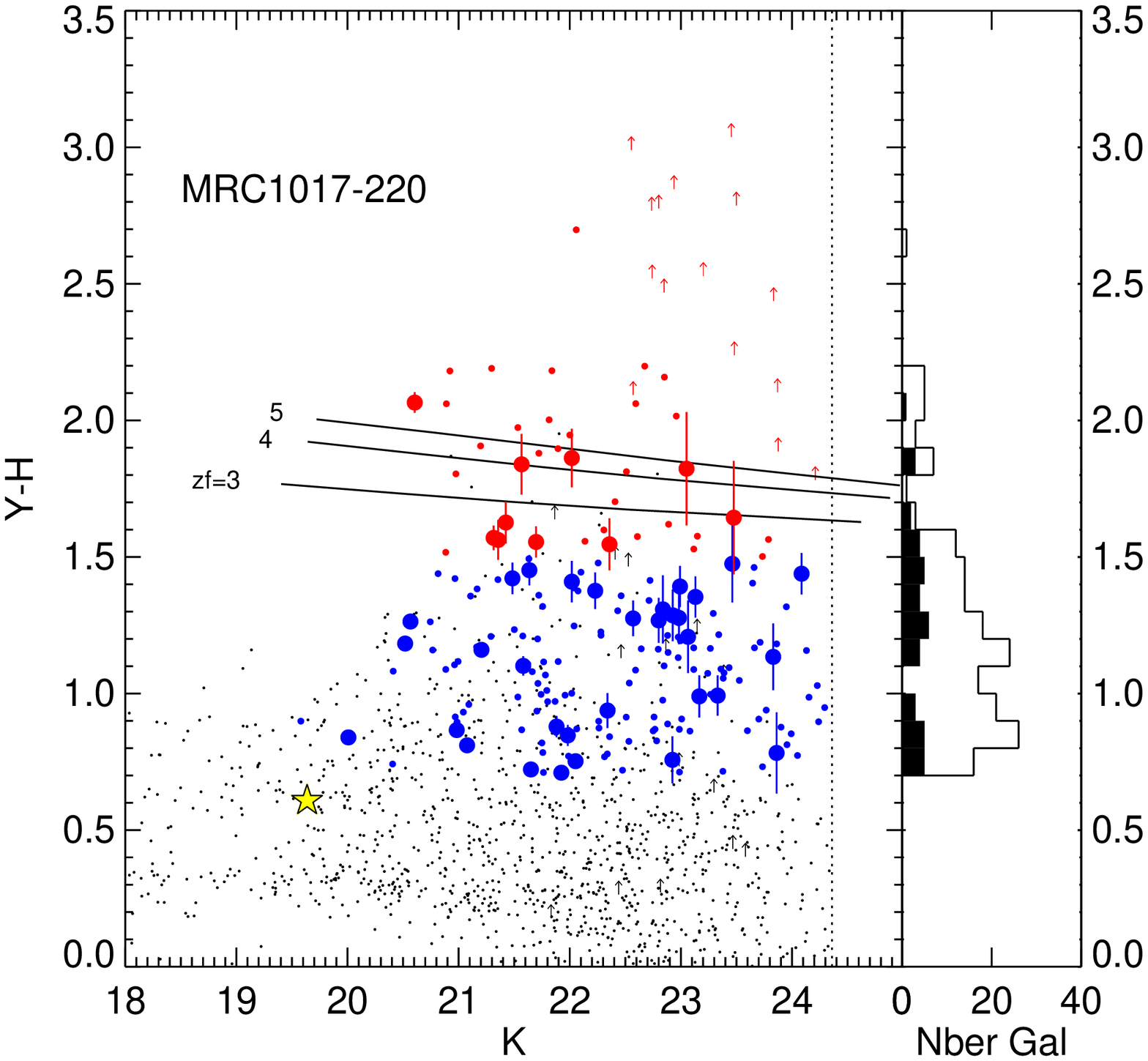} 
\includegraphics[width=8cm,angle=0]{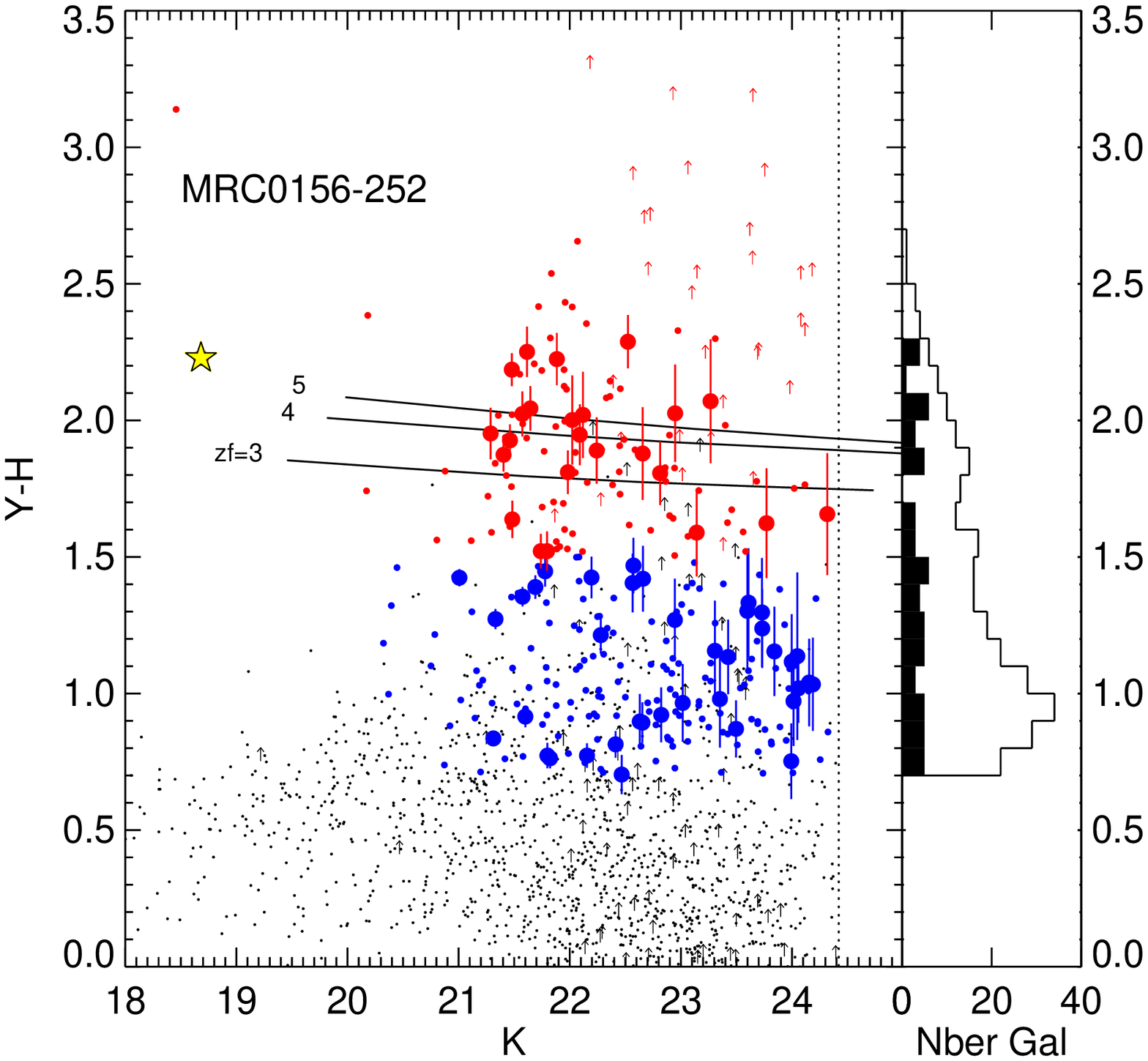} 
\end{center}
\caption{Colour-magnitude diagrams ($Y-H$ vs $K$) of the two GOODS fields and the 
environments of the HzRGs. The r-$YHK$ and b-$YHK$ galaxies are marked by red 
and blue dots. $YHK$ galaxies within $1$~Mpc of the HzRGs are highlighted by bigger symbols 
and error bars in the two lower panels. 
The yellow stars indicate the radio galaxies. The $2\sigma$ detection limit of the $Ks$ bands are 
shown by the dotted lines. Models of a red sequence at $z=1.77$ and $z=2.02$ are 
overplotted in the bottom left and right panels respectively (solid lines). For information, models of red
sequence at $z=1.6$ are also overplotted in the GOODS panels --- corresponding to the redshift of 
the galaxy structure confirmed by Kurk et al.~(2009) in the GOODS field (dashed lines).
Models indicate the predicted colours of a passively evolving stellar population ($z_f=3$, $4$ and $5$ from bottom to top).
We also show on the right side of each plot, the histogram of the $Y-H$ colours of all the candidate 
cluster members (both red and blue galaxies; open histograms). The filled histograms (two lower panels) 
correspond to $YHK$-selected galaxies within $1$~Mpc of the HzRGs.}
\label{cmd}
\end{figure*}

A visual inspection of the field around MRC~0156-252 confirms the 
overdensity of red objects around the HzRG, with the r-$YHK$ being more concentrated 
around the HzRG.
A concentration of b-$YHK$ galaxies is seen in the immediate ($r<2\arcmin$) North-West 
of the HzRG. A second concentration of blue galaxies is also observed in the South-East
part of the field (at R.A.$\sim$29.66-29.70, Dec$\sim$-25.05). However, no counterpart of this 
blue galaxy excess is seen in the r-$YHK$ galaxy distribution.

\subsection{The colour-magnitude diagram}

The distribution of the candidates in the $Y-H$ vs $Ks$ colour-magnitude diagram is shown
in Fig.~\ref{cmd}. Red and blue filled circles indicate r-$YHK$ and b-$YHK$ galaxies respectively. 
Candidates in the close vicinity of the HzRGs ($r<2\arcmin$ corresponding to $1$~Mpc at the HzRG redshifts) 
are shown by larger circles in the bottom panels. 
Both HzRGs are marked by yellow stars\footnote[7]{We note that at the redshift of MRC~1017-220 ($z=1.77$),
the [OII]$\lambda3727$\AA~line is falling in the $Y$ band which can explain the very blue $Y-H$ 
colour of the HzRG.}. We overplot the expected location of a sequence of passively evolving 
galaxies (taken from the Coma cluster at $z=0$) at $z=1.77$ and $z=2.02$ for three different formation 
redshifts ($z_f=3$, $4$ and $5$). Models at $z=1.6$ are also overplotted as the GOODS colour-magnitude 
diagrams (top panels) since a galaxy structure at $z=1.6$ is known to lie in these fields (see \S5.1). 

The right insets of each panel give the histogram of the $Y-H$ colours of the candidates galaxies at $z>1.6$.
It is now well known that galaxies in clusters show a strong bimodality in their colour distribution, 
with star-forming non-dusty galaxies being on the bluer side, and passively evolving galaxies on the 
redder side (corresponding to the location of the red sequence) separated by a `green valley' of
intermediate type objects. A bimodal repartition is suggested in the histograms of the two HzRGs
(lower panels) with two peaks on both sides of $Y-H\sim1.5$ corresponding to the separation 
criteria that was designed to isolate passively evolving galaxies from star-forming blue ones. 

Half of the red sources within $1$~Mpc of MRC~0156-252 have colours consistent with red 
sequence-like galaxies at $z=2.02$. The scatter of these sources is quite large (over $0.4$mag), 
the intrinsic scatter being enlarged by errors on the $Y$-band photometry, and contamination 
from background/foreground galaxies. If some of the red galaxies re indeed passively evolving
galaxies at $z\sim2$, they would have been formed at very high redshift with a formation redshift $z_f>3$.

We also note similarities between the colour-magnitude diagram
of MRC~0156-252 and that of PKS~1138-262 presented in 
\citet{Kodama2007}. The r-$YHK$ galaxies that lie on the red sequence models in the field of
MRC~0156-252 have $21<K<23$ with a high fraction of them ($\sim40$\%) lying on the brighter end 
($21<K<21.5$). This is consistent with the field around PKS~1138-262, where the majority 
of the red near-infrared selected sources ($(J-K)_{Vega}>2.3$) which lay on the red sequence 
had $20.5<K<23$ and half of them had $K<21.5$. This suggests  
that both fields contain massive red galaxies.

At $z\sim2$, the $4000$\AA~break enters the $J$-band. $J$-band spectroscopy 
is thus necessary to prove that the red galaxies are indeed associated with MRC~0156-252. 
\citet{Kriek2008} present a near-infrared spectroscopic survey of $36$ $K$-bright galaxies ($K_{Vega}<19.7$)
at $z_{phot}>2$ selected from the MUlti-wavelength Survey by Yale-Chile 
\citep[MUSYC;][]{Gawiser2006, Quadri2007} and derive spectroscopic redshifts for their full sample
($1.6<z_{spec}<2.73$). Such a spectroscopic campaign could be
conducted similarly in the field of MRC~0156-252 since it contains $23$ r-$YHK$ ($22$\% of the r-$YHK$ sample) 
and $18$ b-$YHK$ galaxies ($13$\% of the b-$YHK$ sample) which have $K<21.5$. 

\citet{Doherty2010} describe a spectroscopic campaign in the PKS~1138-262 field. Two DRGs 
were confirmed to lie at the redshift of the radio galaxy (due to the presence of the H$\alpha$ emission 
line in their spectra). One is a dust-obscured star-forming, red, galaxy. The other 
is an evolved galaxy with little on-going star formation. 
These are the first spectroscopically confirmed red galaxies associated with a protocluster at $z>2$.
These results are encouraging for a future near-IR spectroscopic campaign in the field of MRC~0156-252. 
 
\section{Summary}

We develop a new purely near-infrared $YHK$ $2$-colour selection technique
to isolate galaxies at $z>1.6$ and classify them as {\it (i)} passively evolving, 
or dusty star-forming galaxies and {\it (ii)} star-forming dominated galaxies with little or no dust. 
We test the method using the GOODS-South field,
which has been observed in $Y$, $H$ and $Ks$, and for which a large amount of spectroscopic data
is available. GOODS-S contains a structure of galaxies at $z\sim1.6$ ($42$ spectroscopically confirmed 
members so far). Applying the near-infrared criteria to the GOODS-S field, we recover this 
structure confirming the efficiency of our new selection technique.

We target the surroundings of two high redshift radio galaxies, MRC~1017-220 ($z=1.77$) 
and MRC~0156-252 ($z=2.02$) and a control field using VLT/HAWK-I. The field of MRC~1017-220 shows a 
non-homogeneous filamentary-like spatial arrangement of red galaxies, and a slight 
overdensity of blue galaxies. MRC~0156-252 lies in an overdense region of both blue and red galaxies.
This field is $2-4$ times denser than the other targeted fields. 

The red galaxies are clustered around MRC~0156-252 ($< 1$~Mpc). The blue galaxies 
seem to be preferentially distributed 
in two regions; a concentration of b-$YHK$ galaxies is found immediately at the NW of MRC~0156-252 and
another concentration is found in the SE part of the field, $4\arcmin$ ($\sim2$Mpc at $z=2$) away 
from the HzRG. Our study of the close vicinity of MRC~0156-252 suggests that the radio galaxy has close-by
companions with three galaxies found aligned with the 
HzRG (and the radio axis) and within $5\arcsec$. This structure is similar to the  
system found in the vicinity of PKS~1138-262. 

The distribution of the selected red galaxies in a 
colour-magnitude diagram shows that a large fraction of them have colours consistent with red 
sequence-like objects (with $z_f>3$, assuming they contain no dust). The magnitude range 
of these candidates is also similar to the red protocluster members selected in the 
vicinity of PKS~1138-262, with red galaxies 
preferentially lying on the bright end ($K<21.5$). This suggests that the systems associated
with the HzRGs have already form their more massive members. All these results strongly 
suggest that MRC~0156-252 is associated with a galaxy structure at $z=2$ similar to the galaxy 
system associated with PKS1138-262 at $z=2.16$.

\newcommand{\noopsort}[1]{}


\begin{thebibliography}{85}
\expandafter\ifx\csname natexlab\endcsname\relax\def\natexlab#1{#1}\fi

\bibitem[{{Andreon} {et~al.}(2008){Andreon}, {Maughan}, {Trinchieri}, \&
  {Kurk}}]{Andreon2008}
{Andreon}, S., {Maughan}, B., {Trinchieri}, G., \& {Kurk}, J. 2008,
  ArXiv:0812.1699

\bibitem[{{Balogh} {et~al.}(2007)}]{Balogh2007}
{Balogh}, M.~L. {et~al.} 2007, \mnras, 374, 1169

\bibitem[{{Bertin} \& {Arnouts}(1996)}]{Bertin1996}
{Bertin}, E. \& {Arnouts}, S. 1996, \aaps, 117, 393

\bibitem[{{Best} {et~al.}(2003)}]{Best2003}
{Best}, P.~N. {et~al.} 2003, \mnras, 343, 1

\bibitem[{{Brown} {et~al.}(2003)}]{Brown2003}
{Brown}, M.~J.~I. {et~al.} 2003, \apj, 597, 225

\bibitem[{{Bruzual} \& {Charlot}(2003)}]{Bruzual2003}
{Bruzual}, G. \& {Charlot}, S. 2003, \mnras, 344, 1000

\bibitem[{{Cardelli} {et~al.}(1989){Cardelli}, {Clayton}, \&
  {Mathis}}]{Cardelli1989}
{Cardelli}, J.~A., {Clayton}, G.~C., \& {Mathis}, J.~S. 1989, \apj, 345, 245

\bibitem[{{Casali} {et~al.}(2006)}]{Casali2006}
{Casali}, M. {et~al.} 2006, in SPIE, Vol. 6269, Society of Photo-Optical
  Instrumentation Engineers (SPIE) Conference Series

\bibitem[{{Castellano} {et~al.}(2007)}]{Castellano2007}
{Castellano}, M. {et~al.} 2007, \apj, 671, 1497

\bibitem[{{Castellano} {et~al.}(2010)}]{Castellano2010}
---. 2010, \aap, 511, A260000+

\bibitem[{{Chabrier}(2003)}]{Chabrier2003}
{Chabrier}, G. 2003, \pasp, 115, 763

\bibitem[{{Chen} {et~al.}(2002)}]{Chen2002}
{Chen}, H.-W. {et~al.} 2002, \apj, 570, 54

\bibitem[{{Cimatti} {et~al.}(1999)}]{Cimatti1999}
{Cimatti}, A. {et~al.} 1999, \aap, 352, L45

\bibitem[{{Cimatti} {et~al.}(2000)}]{Cimatti2000}
---. 2000, \mnras, 318, 453

\bibitem[{{Cimatti} {et~al.}(2008)}]{Cimatti2008}
---. 2008, \aap, 482, 21

\bibitem[{{Daddi} {et~al.}(2000)}]{Daddi2000}
{Daddi}, E. {et~al.} 2000, \aap, 361, 535

\bibitem[{{Dickinson} {et~al.}(2003){Dickinson}, {Giavalisco}, \& {GOODS
  Team}}]{Dickinson2003}
{Dickinson}, M., {Giavalisco}, M., \& {GOODS Team}. 2003, in The Mass of
  Galaxies at Low and High Redshift, ed. {R.~Bender \& A.~Renzini}, 324--+

\bibitem[{{Doherty} {et~al.}(2010)}]{Doherty2010}
{Doherty}, M. {et~al.} 2010, \aap, 509, 83

\bibitem[{{Dressler}(1980)}]{Dressler1980}
{Dressler}, A. 1980, \apj, 236, 351

\bibitem[{{Eales} \& {Rawlings}(1996)}]{Eales1996}
{Eales}, S.~A. \& {Rawlings}, S. 1996, \apj, 460, 68

\bibitem[{{Elston} {et~al.}(2006)}]{Elston2006}
{Elston}, R.~J. {et~al.} 2006, \apj, 639, 816

\bibitem[{{Finger} {et~al.}(2008)}]{Finger2008}
{Finger}, G. {et~al.} 2008, in SPIE, Vol. 7021, Society of Photo-Optical
  Instrumentation Engineers Conference Series

\bibitem[{{Franx} {et~al.}(2003)}]{Franx2003}
{Franx}, M. {et~al.} 2003, \apjl, 587, L79

\bibitem[{{Galametz} {et~al.}(2009)}]{Galametz2009B}
{Galametz}, A. {et~al.} 2009, \aap, 507, 131

\bibitem[{{Galametz} {et~al.}(2010)}]{Galametz2010}
---. 2010, ArXiv:1004.3021

\bibitem[{{Gawiser} {et~al.}(2006)}]{Gawiser2006}
{Gawiser}, E. {et~al.} 2006, \apjs, 162, 1

\bibitem[{{Gehrels}(1986)}]{Gehrels1986}
{Gehrels}, N. 1986, \apj, 303, 336

\bibitem[{{Gladders} \& {Yee}(2000)}]{Gladders2000}
{Gladders}, M.~D. \& {Yee}, H.~K.~C. 2000, \aj, 120, 2148

\bibitem[{{Grazian} {et~al.}(2006{\natexlab{a}})}]{Grazian2006A}
{Grazian}, A. {et~al.} 2006{\natexlab{a}}, \aap, 453, 507

\bibitem[{{Grazian} {et~al.}(2006{\natexlab{b}})}]{Grazian2006B}
---. 2006{\natexlab{b}}, \aap, 449, 951

\bibitem[{{Hawarden} {et~al.}(2001)}]{Hawarden2001}
{Hawarden}, T.~G. {et~al.} 2001, \mnras, 325, 563

\bibitem[{{Hewett} {et~al.}(2006)}]{Hewett2006}
{Hewett}, P.~C. {et~al.} 2006, \mnras, 367, 454

\bibitem[{{Hilton} {et~al.}(2007)}]{Hilton2007}
{Hilton}, M. {et~al.} 2007, \apj, 670, 1000

\bibitem[{{Huang} {et~al.}(2001)}]{Huang2001}
{Huang}, J.-S. {et~al.} 2001, \aap, 368, 787

\bibitem[{{Imai} {et~al.}(2007)}]{Imai2007}
{Imai}, K. {et~al.} 2007, \aj, 133, 2418

\bibitem[{{Kajisawa} {et~al.}(2006)}]{Kajisawa2006}
{Kajisawa}, M. {et~al.} 2006, \mnras, 371, 577

\bibitem[{{Kapahi} {et~al.}(1998)}]{Kapahi1998}
{Kapahi}, V.~K. {et~al.} 1998, \apjs, 118, 275

\bibitem[{{Kissler-Patig} {et~al.}(2008)}]{Kissler2008}
{Kissler-Patig}, M. {et~al.} 2008, \aap, 491, 941

\bibitem[{{Kodama} {et~al.}(2007)}]{Kodama2007}
{Kodama}, T. {et~al.} 2007, \mnras, 377, 1717

\bibitem[{{Kong} {et~al.}(2006)}]{Kong2006}
{Kong}, X. {et~al.} 2006, \apj, 638, 72

\bibitem[{{Kriek} {et~al.}(2008)}]{Kriek2008}
{Kriek}, M. {et~al.} 2008, \apj, 677, 219

\bibitem[{{K{\"u}mmel} \& {Wagner}(2001)}]{Kummel2001}
{K{\"u}mmel}, M.~W. \& {Wagner}, S.~J. 2001, \aap, 370, 384

\bibitem[{{Kurk} {et~al.}(2009)}]{Kurk2009}
{Kurk}, J. {et~al.} 2009, ArXiv:0906.4489

\bibitem[{{Kurk} {et~al.}(2004)}]{Kurk2004A}
{Kurk}, J.~D. {et~al.} 2004, \aap, 428, 817

\bibitem[{{Kurk} {et~al.}(2008)}]{Kurk2008A}
{Kurk}, J.~D. {et~al.} 2008, in Astronomical Society of the Pacific Conference
  Series, Vol. 381, Infrared Diagnostics of Galaxy Evolution, 303

\bibitem[{{Lidman} {et~al.}(2008)}]{Lidman2008}
{Lidman}, C. {et~al.} 2008, \aap, 489, 981

\bibitem[{{Maihara} {et~al.}(2001)}]{Maihara2001}
{Maihara}, T. {et~al.} 2001, \pasj, 53, 25

\bibitem[{{McCarthy} {et~al.}(1992){McCarthy}, {Persson}, \&
  {West}}]{McCarthy1992}
{McCarthy}, P.~J., {Persson}, S.~E., \& {West}, S.~C. 1992, \apj, 386, 52

\bibitem[{{Mei} {et~al.}(2006)}]{Mei2006}
{Mei}, S. {et~al.} 2006, \apj, 644, 759

\bibitem[{{Mei} {et~al.}(2009)}]{Mei2009}
---. 2009, \apj, 690, 42

\bibitem[{{Metcalfe} {et~al.}(2006)}]{Metcalfe2006}
{Metcalfe}, N. {et~al.} 2006, \mnras, 370, 1257

\bibitem[{{Miley} \& {De Breuck}(2008)}]{Miley2008}
{Miley}, G. \& {De Breuck}, C. 2008, \aapr, 15, 67

\bibitem[{{Miley} {et~al.}(2004)}]{Miley2004}
{Miley}, G.~K. {et~al.} 2004, \nat, 427, 47

\bibitem[{{Miley} {et~al.}(2006)}]{Miley2006}
---. 2006, \apjl, 650, L29

\bibitem[{{Monet} {et~al.}(2003)}]{Monet2003}
{Monet}, D.~G. {et~al.} 2003, \aj, 125, 984

\bibitem[{{Moy} {et~al.}(2003)}]{Moy2003}
{Moy}, E. {et~al.} 2003, \aap, 403, 493

\bibitem[{{Overzier} {et~al.}(2008)}]{Overzier2008}
{Overzier}, R.~A. {et~al.} 2008, \apj, 673, 143

\bibitem[{{Papovich} {et~al.}(2010)}]{Papovich2010}
{Papovich}, C. {et~al.} 2010, \apj, 716, 1503

\bibitem[{{Pentericci} {et~al.}(2000)}]{Pentericci2000}
{Pentericci}, L. {et~al.} 2000, \aap, 361, L25

\bibitem[{{Pentericci} {et~al.}(2001)}]{Pentericci2001}
---. 2001, \apjs, 135, 63

\bibitem[{{Pickles}(1998)}]{Pickles1998}
{Pickles}, A.~J. 1998, \pasp, 110, 863

\bibitem[{{Pirard} {et~al.}(2004)}]{Pirard2004}
{Pirard}, J. {et~al.} 2004, in The Society of Photo-Optical Instrumentation
  Engineers (SPIE) Conference Series, Vol. 5492, Society of Photo-Optical
  Instrumentation Engineers (SPIE) Conference Series, ed. A.~F.~M. {Moorwood}
  \& M.~{Iye}, 1763--1772

\bibitem[{{Poggianti} {et~al.}(2009)}]{Poggianti2009}
{Poggianti}, B.~M. {et~al.} 2009, \apjl, 697, L137

\bibitem[{{Postman} {et~al.}(2005)}]{Postman2005}
{Postman}, M. {et~al.} 2005, \apj, 623, 721

\bibitem[{{Quadri} {et~al.}(2007)}]{Quadri2007}
{Quadri}, R. {et~al.} 2007, \aj, 134, 1103

\bibitem[{{Rettura} {et~al.}(2010)}]{Rettura2010}
{Rettura}, A. {et~al.} 2010, \apj, 709, 512

\bibitem[{{Retzlaff} {et~al.}(2010)}]{Retzlaff2010}
{Retzlaff}, J. {et~al.} 2010, \aap, 511, A50+

\bibitem[{{Rocca-Volmerange} {et~al.}(2004){Rocca-Volmerange}, {Le Borgne}, {De
  Breuck}, {Fioc}, \& {Moy}}]{Rocca2004}
{Rocca-Volmerange}, B., {Le Borgne}, D., {De Breuck}, C., {Fioc}, M., \& {Moy},
  E. 2004, \aap, 415, 931

\bibitem[{{Rosati} {et~al.}(2004)}]{Rosati2004}
{Rosati}, P. {et~al.} 2004, \aj, 127, 230

\bibitem[{{Rosati} {et~al.}(2009)}]{Rosati2009}
---. 2009, ArXiv:0910.1716

\bibitem[{{Santini} {et~al.}(2009)}]{Santini2009}
{Santini}, P. {et~al.} 2009, ArXiv:0905.0683

\bibitem[{{Saracco} {et~al.}(2001)}]{Saracco2001}
{Saracco}, P. {et~al.} 2001, \aap, 375, 1

\bibitem[{{Schlegel} {et~al.}(1998){Schlegel}, {Finkbeiner}, \&
  {Davis}}]{Schlegel1998}
{Schlegel}, D.~J., {Finkbeiner}, D.~P., \& {Davis}, M. 1998, \apj, 500, 525

\bibitem[{{Seymour} {et~al.}(2007)}]{Seymour2007}
{Seymour}, N. {et~al.} 2007, \apjs, 171, 353

\bibitem[{{Skrutskie} {et~al.}(2006)}]{Skrutskie2006}
{Skrutskie}, M.~F. {et~al.} 2006, \aj, 131, 1163

\bibitem[{{Stanford} {et~al.}(2005)}]{Stanford2005}
{Stanford}, S.~A. {et~al.} 2005, \apjl, 634, L129

\bibitem[{{Stanford} {et~al.}(2006)}]{Stanford2006}
---. 2006, \apjl, 646, L13

\bibitem[{{Tanaka} {et~al.}(2010){Tanaka}, {Finoguenov}, \&
  {Ueda}}]{Tanaka2010}
{Tanaka}, M., {Finoguenov}, A., \& {Ueda}, Y. 2010, ArXiv:1004.3606

\bibitem[{{Tanaka} {et~al.}(2005)}]{Tanaka2005}
{Tanaka}, M. {et~al.} 2005, \mnras, 362, 268

\bibitem[{Vandame(2004)}]{Vandame2004}
Vandame, B. 2004, PhD Thesis, Nice University, France

\bibitem[{{Venemans} {et~al.}(2002)}]{Venemans2002}
{Venemans}, B.~P. {et~al.} 2002, \apjl, 569, L11

\bibitem[{{Venemans} {et~al.}(2005)}]{Venemans2005}
---. 2005, \aap, 431, 793

\bibitem[{{Venemans} {et~al.}(2007)}]{Venemans2007}
---. 2007, \aap, 461, 823

\bibitem[{{Yan} {et~al.}(1998)}]{Yan1998}
{Yan}, L. {et~al.} 1998, \apjl, 503, L19+

\bibitem[{{Zirm} {et~al.}(2008)}]{Zirm2008}
{Zirm}, A.~W. {et~al.} 2008, \apj, 680, 224

\end{thebibliography}
\end{document}